\documentclass[journal]{IEEEtran}
\ifCLASSINFOpdf
\else
\fi
\usepackage{cite}
\usepackage{amsmath,amssymb,amsfonts,mathrsfs,mathtools}
\usepackage{algorithmic}
\usepackage{pict2e}
\usepackage{array}
\usepackage{interval}
\usepackage{hyperref}
\usepackage{color}

\hypersetup{hidelinks}

\def\gof{\mbox{$\boldsymbol{P}\,\#\,\boldsymbol{C}$}}
\newcommand\jw{j\omega}
\newcommand\bbkt[1]{\left\{#1\right\}}
\newcommand\sbkt[1]{\left[#1\right]}
\newcommand\rbkt[1]{\left(#1\right)}
\newcommand\ininf[2]{\langle #1\,, #2 \rangle}

\newcommand\lt{\mathcal{L}_2}
\newcommand\ltp{\mathcal{L}_{2}[0, \infty)}
\newcommand\ltep{\mathcal{L}_{2e}[0, \infty)}
\newcommand\ltc{\mathcal{L}_{2}^\mathbb{C}}
\newcommand\hinf{\mathcal{H}_{\infty}}
\newcommand\linf{\mathcal{L}_{\infty}}

\newcommand\rhinf{\mathcal{RH}_{\infty}}
\newcommand\ccp{\bar{\mathbb{C}}_+}
\newcommand\cn{{\mathbb{C}}^n}
\newcommand\rn{{\mathbb{R}}^n}

\newcommand\sysp{\boldsymbol{P}}

\newcommand\sysi{\boldsymbol{I}}
\newcommand\sysn{\boldsymbol{N}}
\newcommand\mulpi{\boldsymbol{\Pi}}
\newcommand\sysc{\boldsymbol{C}}
\newcommand\sysh{\boldsymbol{H}}
\newcommand\sysm{\boldsymbol{M}}
\newcommand\syspu{\boldsymbol{P}u}
\newcommand\abs[1]{\left|#1\right|}
\newcommand\rep{\text{{R}e}}
\newcommand\imp{\text{{I}m}}
\newcommand{\norm}[1]{\left\lVert#1\right\rVert}

\newcommand{\tbt}[4]{\begin{bmatrix}#1&#2\\#3&#4\end{bmatrix}}

\newcommand{\obt}[2]{\begin{bmatrix}#1&#2\end{bmatrix}}
\newcommand{\tbo}[2]{\begin{bmatrix}#1\\#2\end{bmatrix}}

\newcommand{\bi}{\begin{itemize}}\newcommand{\ei}{\end{itemize}}
\newcommand{\be}{\begin{equation}}\newcommand{\ee}{\end{equation}}
\newcommand{\bex}{\begin{equation*}}\newcommand{\eex}{\end{equation*}}
\newcommand{\bee}{\begin{enumerate}}\newcommand{\eee}{\end{enumerate}}
\newcommand{\bea}{\begin{eqnarray}}\newcommand{\eea}{\end{eqnarray}}
\newcommand{\beas}{\begin{eqnarray*}}\newcommand{\eeas}{\end{eqnarray*}}
\newcommand{\bc}{\begin{center}}\newcommand{\ec}{\end{center}}

\newtheorem{theorem}{Theorem}
\newtheorem{proposition}{Proposition}
\newtheorem{definition}{Definition}
\newtheorem{lemma}{Lemma}
\newtheorem{example}{Example}
\newtheorem{corollary}{Corollary}
\intervalconfig{soft open fences}

\begin{document}
\title{Phase of Nonlinear Systems}
\author{ Chao~Chen,
        Di~Zhao,
        Wei~Chen,
        Sei~Zhen~Khong
        and~Li~Qiu,~\IEEEmembership{Fellow,~IEEE}
\thanks{This work was supported in part by the Guangdong Science and Technology Department, China, under the Grant No.~2019B010117002, the Research Grants
Council of Hong Kong S.A.R, China, under the General Research
Fund No.~16200619, and the National Natural Science Foundation of China under the Grant No.~62073003.}
\thanks{C. Chen and L. Qiu are with the Department of Electronic and Computer Engineering, The Hong Kong University of Science and Technology, Clear Water Bay, Kowloon, Hong Kong S.A.R., China \& The Hong Kong University of Science and Technology Shenzhen Research Institute, Shenzhen 518063, China. E-mails: cchenap@connect.ust.hk, eeqiu@ust.hk. }
\thanks{D. Zhao is with the Department of Control Science and Engineering \& Shanghai Institute of Intelligent Science and Technology, Tongji University, Shanghai 200092, China. E-mail:  dzhao925@tongji.edu.cn.}
\thanks{W. Chen is with the Department of Mechanics and Engineering Science \&
Beijing Innovation Center for Engineering Science and Advanced Technology,
Peking University, Beijing 100871, China. E-mail: w.chen@pku.edu.cn.}
\thanks{S.~Z.~Khong is an independent scholar. {E-mail: szkhongwork@gmail.com}.}
}

\maketitle

\begin{abstract}
In this paper, we propose a definition of phase for a class of stable nonlinear systems called semi-sectorial systems, from an input-output perspective. The definition involves the Hilbert transform as a critical instrument to complexify real-valued signals since the notion of phase arises most naturally in the complex domain. The proposed nonlinear system phase, serving as a counterpart of $\lt$-gain, quantifies the passivity and is highly related to the dissipativity. It also possesses a nice physical interpretation which quantifies the tradeoff between the real energy and reactive energy. A nonlinear small phase theorem is then established for feedback stability analysis of semi-sectorial systems.  Additionally, its generalized version is proposed via the use of multipliers. These nonlinear small phase theorems generalize a version of the classical passivity theorem and a recently appeared linear time-invariant small phase theorem.
\end{abstract}

\begin{IEEEkeywords}
 Small phase theorem, nonlinear systems, passivity, dissipativity, multipliers, Hilbert transform.
\end{IEEEkeywords}

\IEEEpeerreviewmaketitle

\section{Introduction}
\IEEEPARstart{I}{n} the classical frequency-domain analysis of single-input single-output (SISO) linear time-invariant (LTI) systems, the gain and phase go hand in hand, and they are treated on an equal footing in many applications. For more general systems, the gain theory is rich and well established, while its phase counterpart is somewhat inadequate and ambiguous. It is, thus, natural to wonder what a suitable phase definition is for those systems beyond SISO LTI systems. The attempt to answer this question dates back decades, with several notable efforts in exploring phasic information for multi-input multi-output (MIMO) LTI systems, such as the principal phase \cite{Postlethwaite:81} and phase uncertainties \cite{Owens:84, Tits:99, Laib:17}. Recently, a suitable definition of MIMO LTI system phase was proposed in \cite{Chen:19} on the basis of the numerical range. The authors further formulated an LTI small phase theorem for feedback stability analysis which provides a stability condition in terms of the ``loop phase'' being less than $\pi$. We refer the reader to \cite{Chen:19, Chen:21} for more details of the MIMO LTI system phase.

For nonlinear systems, $\lt$-gain is a fundamental quantity used in stability analysis and control of feedback systems from an input-output perspective. The classical small gain theorem \cite{Zames:66} conveys that a feedback system maintains stability provided that its ``loop $\lt$-gain'' is less than one. However, the notion of nonlinear system phase is not well understood. Passivity has been considered as a phase-type counterpart of the $\lt$-gain for a long time. We refer the reader to the book \cite{Van:17} for a comprehensive look at the $\lt$-gain and passivity. A SISO LTI passive system, as is well known, provides a phase-shift of an input sinusoid of at most $\pi/2$. Roughly speaking, this passive system has a ``phase'' within $\interval{-\pi/2}{\pi/2}$. The passivity theorem \cite{Zames:66, Vidyasagar:93, Van:17} ensures the stability of feedback interconnected passive and strictly passive systems, and it is thereupon treated as a ``small phase theorem'' by some researchers \cite{Anders:19}. Another important class of nonlinear systems from a phasic point of view is the counterclockwise systems \cite{Angeli:06}. Briefly, a counterclockwise (or negative imaginary \cite{Petersen:10}) SISO LTI system has a ``phase'' within $\interval{-\pi}{0}$. In conclusion, the passivity and counterclockwise dynamics are only qualitatively phase-related.

The main purpose of this paper is to explore the notion of nonlinear system phase and utilize this quantity in stability analysis of feedback systems.  A major obstacle in defining such a notion is that a practical nonlinear system operates on real-valued signals, while phase is a complex number concept. To overcome this obstacle, we have to artificially introduce complex elements to the real-world systems. This can be achieved by using the Hilbert transform \cite{King:09,Hahn:96}, which plays a vital role in complexifying real-valued signals.

In this paper, we first define the phase for a class of stable nonlinear systems called semi-sectorial systems hereinafter. Second, a nonlinear small phase theorem is established for stability analysis of feedback semi-sectorial systems.  Subsequently, the nonlinear system phase definition and small phase theorem are extended using suitable multipliers. Two essential tools are utilized in the definitions, namely, the numerical range and aforementioned Hilbert transform. The use of these tools in the phase study is inherited from our previous works, in which we proposed the MIMO LTI system phase \cite{Chen:19, Mao:20} and matrix phase \cite{Wang:20}. The nonlinear system phase itself has a nice physical interpretation. In brief, it manifests the tradeoff between the real energy and reactive energy in a signal.  Additionally, our approach to defining the phase is related to the fractional Hilbert transform \cite{Lohmann:96, Lohmann:97, Davis:98, Venkitaraman:14} that has demonstrated its advantages in image processing and optics.

 The nonlinear system phase generalizes the MIMO LTI system phase defined in \cite{Chen:19, Chen:21}. It also admits a strong connection with the static nonlinearity \cite{Vidyasagar:93, Khalil:02}, passivity, counterclockwise dynamics, dissipativity \cite{Willems:72,Hill:80} and integral quadratic constraints (IQCs) \cite{Megretski:97}. In short, the phase of a causal stable passive system is contained in $\interval{-\pi/2}{\pi/2}$; the phases of the static nonlinearity and very strictly passive system can be further well estimated, respectively; the phase-bounded systems can be depicted using the dynamic supply rate or IQC. The nonlinear small phase theorem offers a feedback stability condition involving a comparison between the loop phase and $\pi$. This theorem specializes to the results in \cite{Chen:19, Chen:21} when the open-loop systems are stable and LTI. This theorem is further generalized by virtue of the use of multipliers. By doing so, we extend the applicability of the theorem beyond semi-sectorial systems, and recover the frequency-wise results in \cite{Chen:19, Chen:21} for LTI systems.

The proposed theorem also extends the passivity theorem when the open-loop systems are causal and stable. Specifically, the passivity theorem is conservative in the sense of requiring the phases of open-loop systems to be within $\interval{-\pi/2}{\pi/2}$. In the literature, one common practice to reduce conservatism of the passivity theorem is to quantify passivity using input/output passivity indices \cite{Cho:68, Vidyasagar:77}. These indices, used to measure the surplus or deficit of passivity, can be either positive or negative. Compared with this kind of characterization, the proposed phase offers an alternative quantity from a phasic perspective. Another well-known practice is to adopt the multiplier theorem \cite{Zames:68} by finding a suitable multiplier. The multiplier here, often a noncausal artificial operator, is required to meet a factorization condition which gives rise to difficulties in practice. Generally speaking, the proposed theorem provides an implicit multiplier that is more straightforward and intuitive from a phasic point of view.

The outline of this paper is as follows. In Section~\ref{sec:02}, the preliminaries of signals, systems, the Hilbert transform and analytic signals are introduced. In Section~\ref{sec:03}, we define the phase of nonlinear semi-sectorial systems, and present results and typical examples of the proposed phase. Section~\ref{sec:04} is dedicated to developing nonlinear small phase theorems for feedback stability analysis. Afterwards, a geometric connection is built between the theorems and celebrated circle criterion for the Lur'e systems. Section~\ref{sec:Parallel} presents the phases of parallel and feedback interconnected systems. Section~\ref{sec:05} provides a supplementary discussion on the relation between the phase, dissipativity, IQC and multiplier theorem. Section~\ref{sec:simu} gives simulation results, and Section~\ref{sec:07} concludes this paper.

\section{Preliminaries and Motivations}\label{sec:02}
Let $\mathbb{F}=\mathbb{R}$ or $\mathbb{C}$ be the field of real or complex numbers, and $\mathbb{F}^n$ be the linear space of $n$-dimensional vectors over $\mathbb{F}$. Denote ${\mathbb{R}}_{+}$ and $\ccp$ as  the set of positive real numbers and the closed complex right half-plane, respectively. The conjugate, transpose and conjugate transpose are denoted by$\overline{(\cdot)}$, $(\cdot)^T$ and $(\cdot)^*$, respectively. For $x\in \mathbb{F}^n$, denote $|x|$ as its Euclidean norm. The real and imaginary parts of a complex number $z\in \mathbb{C}$ are denoted by $\rep\rbkt{z}$ and $\imp\rbkt{z}$, respectively. The angle of a nonzero $z\in \mathbb{C}$ in the polar form $\abs{z}e^{j\angle z}$ is denoted by $\angle z$. If $z=0$, then $\angle z$ is undefined. The Hermitian part of a square matrix $A\in \mathbb{C}^{n\times n}$ is denoted by $\text{He}(A)\coloneqq \rbkt{A+A^*}/2$. In addition, $A>0$ ($A\geq 0$, resp.) denotes that $A$ is positive definite (positive semi-definite, resp.). Denote $\rhinf^{n\times n}$ as the space consisting of  $n\times n$ real rational proper matrix-valued functions with no poles in $\ccp$ and $\rhinf\coloneqq \rhinf^{1\times 1}$.  Let $\linf^{n\times n}$  denote the space consisting of $n \times n$ matrix-valued functions which are essentially bounded on the imaginary axis $j{\mathbb{R}}$ and $\linf\coloneqq \linf^{1\times 1}$.

\subsection{Signal Spaces, Operators and Systems}
The input-output analysis of nonlinear systems is often built on a real signal space. We start with the $\lt$ space, the set of all energy-bounded $\rn$-valued signals:
\bex
\lt \coloneqq \left\{ u\colon \mathbb{R}\rightarrow\rn \Big|~\norm{u}_2^2 \coloneqq \int_{-\infty}^{\infty} |u(t)|^2dt < \infty\right\}.
\eex
The causal subspace of $\lt$ is denoted by $\ltp \coloneqq \left\{ u \in \lt |~u(t)=0~\text{for}~t < 0 \right\}$. For $T\in \mathbb{R}$, define the truncation $\boldsymbol{\Gamma}_T$ on all $u\colon \mathbb{R} \rightarrow \rn$ by $(\boldsymbol{\Gamma}_T u)(t)\coloneqq u(t)$ for $t\leq T$;   $(\boldsymbol{\Gamma}_T u)(t)\coloneqq 0$ for $t>T$. The extended space of $\ltp$ is then denoted by $\ltep \coloneqq \left\{ u\colon \mathbb{R} \rightarrow\rn |~\boldsymbol{\Gamma}_T u \in \ltp, \forall T\geq 0 \right\}$.

An operator $\sysp\colon \ltep\to\ltep$ is said to be causal if $\boldsymbol{\Gamma}_T\sysp=\boldsymbol{\Gamma}_T\sysp\boldsymbol{\Gamma}_T$ for all $T\geq 0$, and is said to be noncausal if it is not causal. The causality of an operator $\sysp\colon \lt\to\lt$ is defined in the same manner, except for $T\in \mathbb{R}$. We always assume that an operator $\sysp$ maps the zero signal to the zero signal, i.e., $\sysp0=0$.  We view a system as an operator from real-valued input signals to real-valued output signals. In addition, we consider only ``square'' systems with the same number of inputs and outputs, and assume that these systems are nonzero, i.e., $\sysp\neq 0$.

A practical nonlinear system is represented by a causal operator $\sysp\colon \ltep \rightarrow \ltep$. The $\lt$ domain of $\sysp$, namely, the set of all its input signals in $\ltp$ such that the output signals are in $\ltp$, is denoted by $\text{dom}(\sysp)\coloneqq \left\{ u\in \ltp |~ \syspu \in \ltp  \right\}$.
Such a causal system $\sysp$ (operator, resp.) is said to be stable (bounded, resp.) if $\text{dom}(\sysp) = \ltp$ and
\be\label{eq:l2gain}
 \norm{\sysp} \coloneqq  \sup_{0\neq u \in \ltp} \frac{\norm{\sysp u}_2}{\norm{u}_2} < \infty.\ee
Here, $\norm{\sysp}$ is called the $\lt$-gain of $\sysp$,  and it is the key quantity used in the gain-based input-output nonlinear system control theory. For the causal system $\sysp$, it is well-known that the stability on $\ltp$ is equivalent to the stability on $\ltep$ \cite[Proposition 1.2.3]{Van:17}. Then, in addition to (\ref{eq:l2gain}), it holds that
\bex \norm{\sysp} =\sup_{\substack{u \in \ltep, T>0, \\  \norm{\boldsymbol{\Gamma}_T u}_2\neq 0}} \frac{\norm{\boldsymbol{\Gamma}_T \sysp u}_2}{\norm{\boldsymbol{\Gamma}_T u}_2}.\eex

Passivity is another key notion for input-output analysis of nonlinear systems. A causal stable $\sysp$ is called passive \cite{Van:17} if
\be\label{eq:passive}
\ininf{u}{\sysp u}\geq 0,\quad \forall u \in \ltp.
\ee
Moreover, it is called very strictly passive if there exist $\delta>0$ and $\epsilon>0$ such that \be\label{eq:VSP}
\ininf{u}{\sysp u}\geq \delta \norm{u}^2_2+\epsilon \norm{\sysp u}^2_2,\quad \forall u \in \ltp;
\ee
it is called output strictly passive if (\ref{eq:VSP}) holds with $\delta=0$.

For a SISO LTI passive system $\sysp$ with transfer function $P(s)\in \rhinf$, the condition (\ref{eq:passive}) is equivalent to that $\rep\rbkt{P(\jw)} \geq 0$ for all $\omega \in {\mathbb{R}}$.
Thus, it is understood that $\angle P(\jw)$ lies in $\interval{-\pi/2}{\pi/2}$. We see that the passivity is phase-related, but only qualitatively.

We aim to develop a quantifiable phasic notion of nonlinear systems. To this end, observe an obvious problem that a practical nonlinear system can only accept and generate real-valued signals. However,  phases are naturally defined for complex numbers. For example, phases of SISO LTI systems are introduced in terms of their frequency responses. To overcome this problem, we have to introduce complex elements to real-world systems. Therefore, the fundamental nontrivial question behind a nonlinear system phase definition is how we can appropriately complexify real-valued signals. Our answer is to utilize the Hilbert transform to obtain the corresponding analytic signals, which are complex-valued and commonly used in signal processing. Accordingly, in our study, we will also need the complex signal space $\ltc$ which denotes the set of all energy-bounded $\cn$-valued signals:
\bex
\ltc \coloneqq \left\{ u\colon\mathbb{R}\rightarrow\cn \Big|~\norm{u}_2^2 \coloneqq \int_{-\infty}^{\infty} |u(t)|^2dt < \infty\right\}.
\eex
Let $\hat{u}$ denote the Fourier transform of a signal $u\in \ltc$. By the well-known Plancherel's theorem, for all $u, v\in\ltc$, we have \bex\ininf{u}{v} = \ininf{\hat{u}}{\hat{v}} \coloneqq \frac{1}{2\pi}\int_{-\infty}^{\infty}\hat{u}(\jw)^*\hat{v}(\jw)d\omega.
\eex
Furthermore, the Fourier transform on $\ltc$ is a unitary operator.

\subsection{The Hilbert Transform}
The \textit{Hilbert transform} $\boldsymbol{H}$ of a complex-valued signal $u(t)$ is defined by the integral \cite{King:09}
\bex
(\boldsymbol{H}u)(t) \coloneqq \frac{1}{\pi} \int_{-\infty}^{\infty}\frac{u(\tau)}{t-\tau} d\tau =\frac{1}{\pi t} \ast u(t),
\eex
where $\ast$ denotes the convolution operation, provided that the integral exists. The integral above is improper due to the pole at $\tau=t$ and is evaluated in the sense of the Cauchy principal value. A simple example is the signal $f(t)=\cos(a t)$ whose Hilbert transform is given by
\bex
(\sysh f)(t)= \text{sgn}(a)\cos(at-\frac{\pi}{2})=\text{sgn}(a)\sin(at),
\eex
where $a\in \mathbb{R}$ and $\text{sgn}(\cdot)$ denotes the signum function. This example gives an intuition that, the Hilbert transform provides a pure $\pi/2$ phase-shift, which can be clarified conveniently using the frequency-domain language. Specifically, the Fourier transform of the convolution kernel $1/(\pi t)$ is given by $-j\text{sgn}(\omega)$. This gives
\be\label{eq:freqH}
(\widehat{\boldsymbol{H}u})(j\omega)=-j\text{sgn}(\omega)\hat{u}(j\omega).
\ee
Therefore, the Hilbert transform provides a $-\pi/2$ phase-shift for positive frequencies, while a $\pi/2$ phase-shift for negative frequencies. Concurrently, the magnitudes of the spectrum remain unchanged. For a differentiable signal $u(t)$, it is worthwhile to briefly compare its Hilbert transform with its derivative $\dot{u}(t)$. The Fourier transform of the latter signal $\dot{u}(t)$ is given by $(\widehat{\dot{u}})(j\omega)=j\omega\hat{u}(j\omega)$,
which also offers a $\pi/2$ phase-shift, while its magnitude varies with $\omega$.

In the rest of this paper, we restrict the Hilbert transform to $\ltc$. It is well known that the Hilbert transform $\boldsymbol{H}\colon\ltc \rightarrow \ltc$ defines a noncausal linear bounded operator possessing three favorable properties:
 \bee
  \item isometry: $\norm{\boldsymbol{H}u}_2=\norm{u}_2$;
  \item anti-self-adjointness: $\boldsymbol{H}^*=-\boldsymbol{H}$;
  \item anti-involution: $\boldsymbol{H}(\boldsymbol{H}u)=-u$
\eee
for $u \in \ltc$, where $\sysh^*$ denotes the adjoint operator of $\sysh$. The isometry and anti-self-adjointness can be handily proved using the Plancherel's  theorem, i.e., for all $u,v\in \ltc$, we have
\begin{align*}
\norm{\boldsymbol{H}u}_2^2=\norm{\widehat{\boldsymbol{H}u}}_2^2=&\frac{1}{2\pi}\int_{-\infty}^{\infty}\abs{-j\text{sgn}(\omega)
 {\hat{u}(\jw)}}^2 d\omega\\
 =&\frac{1}{2\pi}\int_{-\infty}^{\infty}\abs{{\hat{u}(\jw)}}^2 d\omega=\norm{\hat{u}}^2_2=\norm{u}^2_2,\\
 \ininf{\sysh u}{v}=\ininf{\widehat{\sysh u}}{\hat{v}}=&\frac{1}{2\pi}\int_{-\infty}^{\infty}\sbkt{-j\text{sgn}(\omega)
 {\hat{u}(\jw)}}^* \hat{v}(\jw) d\omega\\
 =&\ininf{\hat{u}}{\widehat{ -\sysh v}}=\ininf{u}{-\sysh v},
\end{align*}
respectively. The anti-involution follows straightforwardly from the isometry and anti-self-adjointness. To sum up,  the Hilbert transform on $\ltc$ preserves the inner product, and in particular, it is a unitary operator.

The Hilbert transform is a lossless process on account of merely generating a phase-shift to the original signal. Thus, it is often utilized to generate a complex-valued signal from a real-valued signal in signal processing \cite{Gabor:46,Hahn:96}. A complex-valued signal, whose imaginary part is the Hilbert transform of its real part, is called an \textit{analytic signal}. Specifically, the analytic representation of a signal $u \in \ltc$ is denoted by
\be\label{eq:analytic} u_a(t) \coloneqq  \frac{1}{2}(u(t)+ j (\boldsymbol{H}u)(t)).\ee
In connection with the complexification of real-valued signals in nonlinear systems, for $u \in \lt$, the definition in (\ref{eq:analytic}) gives rise to $u_a \in \ltc$.

Several beneficial properties of the analytic signals are elaborated as follows. Firstly, for any real-valued signal $u\in \lt$, it holds that $\ininf{u}{\boldsymbol{H}u}=0$. This can be deduced from
\begin{align*}
 \ininf{u}{\boldsymbol{H}u}=\ininf{\hat{u}}{\widehat{\boldsymbol{H}u}}=
 \frac{1}{2\pi}\int_{-\infty}^{\infty}\sbkt{-j\text{sgn}(\omega)}
 \abs{\hat{u}(\jw)}^2d\omega=0,
\end{align*}
where the first equality is based on the Plancherel's theorem, and the last uses the facts that,  $\text{sgn}(\omega)$ is odd, and $\abs{\hat{u}(\jw)}^2$  is even by conjugate symmetry of  $\hat{u}(\jw)$. This implies that a real-valued signal $u(t)$ and its Hilbert transform $(\boldsymbol{H}u)(t)$ are orthogonal. Accordingly, $(\boldsymbol{H}u)(t)$ is referred to as the quadrature function of $u(t)$ in signal processing.

Secondly, for all real-valued signals $u, v \in \lt$, one can derive the following useful identities:
\be\label{eq:analyticalproperty}
\norm{u_a}_2^2 = \frac{1}{2} \norm{u}_2^2\quad \text{and}\quad \ininf{u_a}{v} = \ininf{u}{v_a} = \ininf{u_a}{v_a}
\ee
from the aforementioned properties of orthogonality, anti-self-adjointness and anti-involution. The above identities will be frequently utilized in the rest of this paper. Equipped with the analytic signals, we are now ready to define the nonlinear system phase from an input-output perspective.

\section{Nonlinear System Phases}\label{sec:03}
To highlight the main idea and for the sake of brevity, we provide all the proofs in this paper in Appendix.
\subsection{The Phase of Nonlinear Systems}
\begin{figure}[htb]
\vspace{-2mm}
\centering
\includegraphics[width=2in]{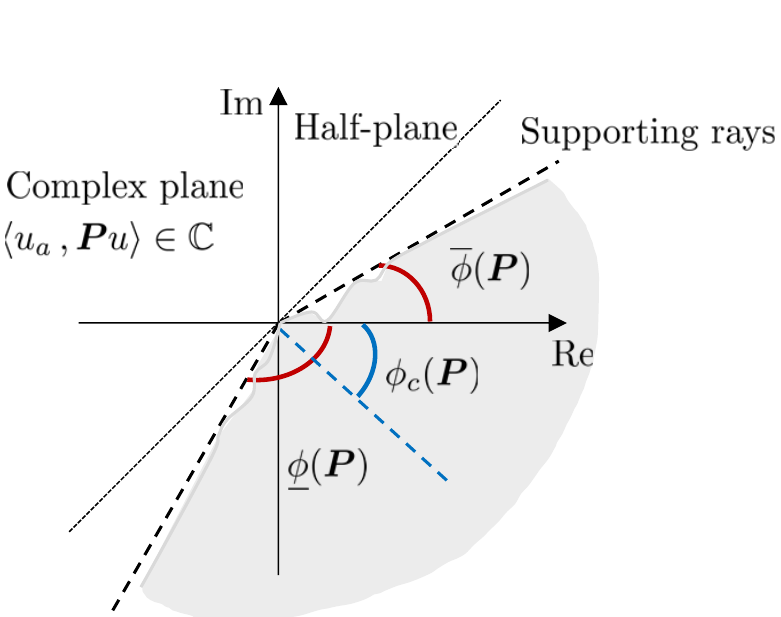}
\caption{An illustration of the angular numerical range $W^{\prime}(\sysp)$ (gray) with its two supporting rays (dashed lines) and the phase
sector ${\Phi}(\sysp) = \interval{\underline{\phi}(\sysp)}{\overline{\phi}(\sysp)}$ (red) of a semi-sectorial nonlinear system $\sysp$.} \label{fig:Nrange}
\end{figure}

Consider a causal stable nonlinear system $\sysp\colon\ltep \rightarrow \ltep$. Based on the notion of analytic signals, the \textit{angular numerical range} of $\sysp$ is defined to be \be \label{eq:nrange}
W^{\prime}(\sysp) \coloneqq \bbkt{ \ininf{u_a}{\syspu} \in \mathbb{C}~|~ u \in \ltp }.  \ee
Such a system $\sysp$ is said to be \textit{semi-sectorial} if $W^{\prime}(\sysp)$ is contained in a closed complex half-plane. Geometrically for a semi-sectorial system $\sysp$, there are two unique supporting rays of $W^{\prime}(\sysp)$ to form an angular sector. See Fig.~\ref{fig:Nrange} for an illustration of the angular numerical range $W^{\prime}(\sysp)$ with its two supporting rays.  Denote $\phi_c(\sysp) \in \interval[open left]{-\pi}{\pi}$ as the angle from the positive real axis to the interior angle bisector of these two rays. Then, the \textit{phase} of $\sysp$, denoted by ${\Phi}(\sysp)$, is defined to be the \textit{phase sector}
\bex\label{eq:phase}
{\Phi}(\sysp) \coloneqq \interval{\underline{\phi}(\sysp)}{\overline{\phi}(\sysp)},
\eex
where ${\underline{\phi}(\sysp)}$ and ${\overline{\phi}(\sysp)}$ are called the phase infimum and phase supremum of $\sysp$, respectively, which are defined by
\begin{align*}
{\underline{\phi}}(\sysp) \coloneqq \inf_{0\neq z \in W^{\prime}(\sysp)} {\angle z}\quad \text{and}\quad
{\overline{\phi}}(\sysp) \coloneqq \sup_{0\neq z \in W^{\prime}(\sysp)} {\angle z}.
\end{align*}
Both ${\underline{\phi}(\sysp)}$ and ${\overline{\phi}(\sysp)}$ take value in $\interval{\phi_c(\sysp)-\pi/2}{\phi_c(\sysp)+\pi/2}$. Moreover, ${\overline{\phi}(\sysp)}-{\underline{\phi}(\sysp)}\in \interval{0}{\pi}$ is called the \textit{phase spread} of $\sysp$.  By definition, we have \bex\phi_c(\sysp)=\frac{1}{2}\rbkt{{\underline{\phi}(\sysp)}+{\overline{\phi}(\sysp)}}.\eex In the rest of this paper, the two terms, the phase and phase sector, will be used interchangeably without ambiguity.

Each $\ininf{u_a}{\syspu} \in W^{\prime}(\sysp)$ in $\mathbb{C}$ is associated with a point $\rbkt{\rep\ininf{u_a}{\syspu},  \imp\ininf{u_a}{\syspu}}$ in $\mathbb{R}^2$.  Accordingly, $\sysp$ is semi-sectorial if and only if there exists a real number $\alpha \in \interval{-\pi}{\pi}$ such that ${\Phi}(\sysp) \subseteq \interval{-{\pi}/{2}-\alpha}{{\pi}/{2}-\alpha}$, which is further equivalent to
\be\label{eq:sectorial}
\cos\alpha\rep\ininf{u_a}{\syspu} - \sin\alpha\imp\ininf{u_a}{\syspu} \geq 0\ee
for all $u \in \ltp$. Geometrically, this inequality means that the set of points $\rbkt{\rep\ininf{u_a}{\syspu},  \imp\ininf{u_a}{\syspu}}$ is contained in a closed half-plane in $\mathbb{R}^2$, with its normal vector being $\obt{\cos\alpha}{-\sin\alpha}^T$. In addition, $\sysp$ is said to be \textit{sectorial} if there exist $\delta, \epsilon\geq 0$ with $\delta+\epsilon>0$ such that
\be\label{eq:ISS}
 \cos\alpha\rep\ininf{u_a}{\syspu} - \sin\alpha\imp\ininf{u_a}{\syspu}  \geq \delta\norm{u}^2_2 +\epsilon\norm{\sysp u}^2_2 \ee
for all $u \in \ltp$. That is to say, every point satisfying inequality (\ref{eq:ISS}) deviates from the origin along the normal vector. This implies that the phase spread ${\overline{\phi}(\sysp)}-{\underline{\phi}(\sysp)}<\pi$. Since $\sysp$ is stable, in light of (\ref{eq:ISS}), we have
\begin{flalign*}
 \cos\alpha\rep\ininf{u_a}{\syspu} - \sin\alpha\imp\ininf{u_a}{\syspu}  \geq \delta\norm{u}^2_2 + \epsilon\norm{\sysp u}^2_2\\
   \geq \rbkt{\frac{\delta}{\norm{\sysp}^2}+\epsilon}\norm{\sysp u}_2^2=\hat{\epsilon}\norm{\sysp u}_2^2,
\end{flalign*}
where $\hat{\epsilon}\coloneqq\delta/\norm{\sysp}^2 +\epsilon> 0$. This shows that, for a sectorial system $\sysp$, (\ref{eq:ISS}) is equivalent to there existing $\hat{\epsilon}>0$ such that
\begin{align}\label{eq:SS}
 \cos\alpha\rep\ininf{u_a}{\syspu} - \sin\alpha\imp\ininf{u_a}{\syspu}  \geq \hat{\epsilon}\norm{\sysp u}_2^2.
\end{align}
Thus, (\ref{eq:SS}) will hereinafter be adopted as the equivalent definition of sectorial systems.

A bridge between the nonlinear system phase and passivity can be evidently built. Recall that, by (\ref{eq:passive}), a causal stable system $\sysp$ is passive if $\ininf{u}{\sysp u}\geq 0$ for all $u\in\ltp$, which is equivalent to $W^{\prime}(\sysp) \subseteq \ccp$ due to the fact that $\ininf{u}{\sysp u}=2\rep\ininf{u_a}{\sysp u} \geq 0$. Therefore, from the definition of phase, such a passive system $\sysp$ has its phase $\Phi{(\sysp)}\subseteq\interval{-\pi/2}{\pi/2}$. This manifests that the passivity is qualitative, while the phase is quantitative. Later in Section~\ref{sec:3b}, we will further show that for some classes of passive systems, we can have a more precise estimation of their phases.

Last but not least, all the aforementioned definitions of nonlinear systems can be generalized via the use of multipliers. Let a multiplier $\mulpi\colon \ltc \rightarrow \ltc$ be a bounded LTI operator with transfer function $\Pi(s) \in \linf^{n\times n}$. The \textit{angular $\mulpi$-numerical range} of $\sysp$ with respect to $\boldsymbol{\Pi}$ is defined to be
\bex
W_{\mulpi}^{\prime}(\sysp) \coloneqq \bbkt{ \ininf{\mulpi u_a}{\syspu} \in \mathbb{C}~|~ u \in \ltp }.  \eex
The standard angular numerical range $W^{\prime}(\sysp)$ can be recovered by taking $\mulpi$ to be the identity operator $\sysi$. The corresponding \textit{$\mulpi$-semi-sectorial} and \textit{$\mulpi$-sectorial} systems with their \textit{$\mulpi$-phases} can be defined in parallel to the standard ones. For example,  based on (\ref{eq:SS}), a $\mulpi$-sectorial system satisfies
\bex
 \cos\alpha\rep\ininf{\mulpi u_a}{\syspu} - \sin\alpha\imp\ininf{\mulpi u_a}{\syspu}  \geq \hat{\epsilon}\norm{\sysp u}^2_2 \eex
for all $u\in \ltp$, with some $\hat{\epsilon} >0$ and $\alpha\in \interval{-\pi}{\pi}$. Hereinafter, we adopt the subscript $\mulpi$ to indicate that the multiplier $\mulpi$ is involved in the definitions. We will elaborate the use of $\mulpi$ in Section~\ref{sec:04}, which links our results to the frequency-wise LTI small phase theorem in \cite{Chen:19, Chen:21}.

\subsection{Examples: The Phase of Typical Nonlinear Systems}\label{sec:3b}
\subsubsection{From a Nonlinearity Sector to a Phase Sector}
The sector bounded static nonlinearity \cite[Section~5.6]{Vidyasagar:93} is a representative nonlinear system which is worthy of consideration using phasic language. This nonlinearity is widely adopted in the modeling of open-loop systems, e.g., the Wiener-Hammerstein systems \cite{Narendra:66}, and in that of closed-loop systems, e.g., the Lur'e systems \cite[Section~5.6]{Vidyasagar:93}. Below, we will show the phase estimation of a sector bounded static nonlinearity.

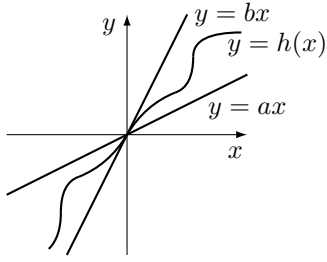
\begin{figure}[htbp]
\centering
\setlength{\unitlength}{0.8mm}
\begin{picture}(40,42)
\put(0,20){\vector(1,0){40}}
\put(20,0){\vector(0,1){40}}
\thicklines
\put(20,20){\line(2,1){20}}
\put(20,20){\line(-2,-1){20}}
\put(20,20){\line(1,2){10}}
\put(20,20){\line(-1,-2){10}}
\qbezier(20, 20)(23, 25)(28, 27)
\qbezier(28, 27)(31, 28)(31, 33)
\qbezier(31, 33)(31, 37)(39, 37)
\qbezier(20, 20)(17, 15)(12, 13)
\qbezier(12, 13)(9, 12)(9, 7)
\qbezier(9, 7)(9, 3)(7, 1)
\put(40,20){\makebox(0,8){$y=a x$}}
\put(35,40){\makebox(5,0){$y=b x$}}
\put(40,35){\makebox(10,0){$y=h(x)$}}
\put(38,17){\makebox(0,0){$x$}}
\put(17,38){\makebox(0,0){$y$}}
\end{picture}\caption{A system $\sysn$ in the nonlinearity sector from $a$ to $b$, with $b>a>0$.} \label{fig:sector}
\end{figure}

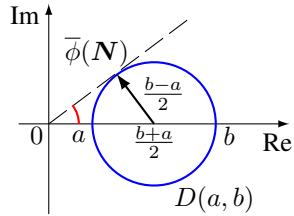
\begin{figure}[htb]
\centering
\setlength{\unitlength}{0.8mm}
\begin{picture}(40,40)
\put(0,20){\vector(1,0){45}}
\put(5,5){\vector(0,1){35}}
\thicklines
\put(3,18){\makebox(0,0){$0$}}
\put(30,5){\makebox(5,5){$D(a, b)$}}
{\color{blue}\put(22.5,20){\circle{20.5}}}
\put(10,18){\makebox(0,0){$a$}}
\put(35,18){\makebox(0,0){$b$}}
\put(22.5,17){\makebox(0,0){$\frac{b+a}{2}$}}
\put(24,25){\makebox(0,0){$\frac{b-a}{2}$}}
\put(13,32){\makebox(0,0){$\overline{\phi}(\sysn)$}}
\thinlines\multiput(5,20)(4,3){6}{\line(4,3){3}}
\thicklines
\put(22.5,20){\vector(-3,4){6.3}}
{\color{red}\put(5, 20){\arc[0,38]{5}}}
\put(43,17){\makebox(0,0){$\rep$}}
\put(1,38){\makebox(0,0){$\imp$}}
\end{picture}\caption{The disk $D(a, b)$ in $\mathbb{C}$ with the center $(b+a)/2+j0$ and radius $(b-a)/2$ when $b>a>0$.}\label{fig:Phase_sector}
\end{figure}

Consider a scalar static nonlinear system $\sysn\colon \ltep \rightarrow \ltep$ defined by
\be\label{eq:staticnonlinear}
(\sysn u)(t)=h(u(t)),
\ee
where $h\colon \mathbb{R}\rightarrow\mathbb{R}$ satisfies
\be\label{eq:sector}
\rbkt{h(x)-a x}\rbkt{h(x) -b x } \leq 0,\quad \forall x \in \mathbb{R}
\ee with $b > a > 0$. The graphical representation of $h$, illustrated by Fig.~\ref{fig:sector},  belongs to a sector bounded by the two lines $y=a x$ and $y=b x$. Consequently, we say $\sysn$ is a sector bounded static nonlinearity, and it belongs to the nonlinearity sector from $a$ to $b$. Clearly, $\sysn$ is passive since $u(t)(\sysn u)(t)$ is nonnegative for every $t\geq 0$. Let $D(a, b)$ denote the closed disk in $\mathbb{C}$, illustrated by Fig.~\ref{fig:Phase_sector}, with the
center $(b+a)/2$ and radius $(b-a)/2$, namely,
   \bex
D (a, b) \coloneqq \bbkt{z\in \mathbb{C} \Bigg|~\abs{z - \frac{b+a}{2}} \leq \frac{b-a}{2}}.
    \eex
The following proposition helps estimate the phase sector $\Phi(\sysn)$ from the nonlinearity sector.

\begin{proposition}\label{lem:sector}
For a scalar system $\boldsymbol{N}$ satisfying the nonlinearity sector condition (\ref{eq:sector}), we have
  \bex \angle \ininf{u_a}{\boldsymbol{N}u} \in \interval{\min \angle D(a, b)}{\max \angle D(a, b)} \eex
 for all $0 \neq u\in \ltp$. A simple computation shows that \be\label{eq:phase_sector}
{\Phi}(\boldsymbol{N})\subseteq\interval{-{\arcsin\frac{b-a}{b+a}}}{\arcsin\frac{b-a}{b+a}},
  \ee
where $\arcsin(\cdot)\colon\interval{-1}{1} \rightarrow \interval{-\pi/2}{\pi/2}$.
\end{proposition}

In Proposition~\ref{lem:sector}, the phase of $\sysn$ is contained in the phase spread of the disk $D(a, b)$ shown in Fig.~\ref{fig:Phase_sector}. This disk $D(a, b)$ is exactly $-ab$ times the ``disk'' in the celebrated circle criterion \cite[Section~5.2]{Desoer:75}, which will be elaborated in Section~\ref{sec:04:a} when the Lur'e problem is studied.

\begin{figure}[htb]
\centering
\setlength{\unitlength}{0.8mm}
\begin{picture}(40,40)
\put(-10,15){\vector(1,0){60}}
\put(20,0){\vector(0,1){35}}
\put(22,13){\makebox(0,0){$0$}}
\multiput(20,15)(4,3){6}{\line(4,3){3}}
\multiput(20,15)(-4,-3){6}{\line(-4,-3){3}}
\multiput(20,15)(5,1){6}{\line(6,1){3}}
\multiput(20,15)(-5,-1){6}{\line(-6,-1){3}}
\put(48,13){\makebox(0,0){$x$}}
\put(15,32){\makebox(0,0){$h(x)$}}
\thicklines
{\color{blue}
\put(19.3,14.8){\line(0,-1){0.2}}
\put(19.3,14.6){\line(-1,0){1}}
\put(18.3,14.6){\line(0,-1){0.6}}
\put(18.3,14){\line(-1,0){2.8}}
\put(15.5,14){\line(0,-1){2.2}}
\put(15.5,11.8){\line(-1,0){11.5}}
\put(4,11.8){\line(0,-1){8.7}}
\put(4,3.1){\line(-1,0){12}}
\put(20.7,15.2){\line(0,1){0.2}}
\put(20.7,15.4){\line(1,0){1}}
\put(21.7,15.4){\line(0,1){0.6}}
\put(21.7,16){\line(1,0){2.8}}
\put(24.5,16){\line(0,1){2.2}}
\put(24.5,18.2){\line(1,0){11.5}}
\put(36,18.2){\line(0,1){8.7}}
\put(36,26.9){\line(1,0){12}}
}
\end{picture}\caption{A logarithmic quantizer with $\rho=1/3$.}\label{fig:Log}
\end{figure}
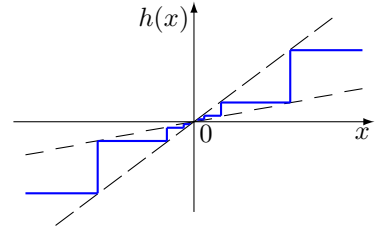

\begin{example}
Consider a logarithmic quantizer $\sysn$ defined by (\ref{eq:staticnonlinear}), with $h\colon \mathbb{R}\rightarrow\mathbb{R}$ satisfying
\bex
h(x) = \left\{
\begin{array}{ll}
  \rho^i, &\text{if}~x \in \interval[open left]{\frac{1+\rho}{2}\rho^i}{ \frac{1+\rho}{2\rho}\rho^i},~i=0, \pm1, \cdots\\
   0,&\text{if}~x=0,\\
   -h(-x),&\text{if}~x<0,
\end{array}
\right.
\eex
where $\rho \in \interval[open]{0}{1}$ represents the quantization density \cite{Fu:05_2}, i.e., a small $\rho$ means a coarse quantizer.  The graphical representation of $h$ is shown in Fig.~\ref{fig:Log}. It is known that $\sysn$ belongs to the nonlinearity sector from ${{2\rho}/({1+\rho})}$ to ${{2}/({1+\rho})}$.  It follows from Proposition~\ref{lem:sector} that \bex
\Phi(\sysn)\subseteq \interval{-\arcsin\frac{1-\rho}{1+\rho}}{\arcsin\frac{1-\rho}{1+\rho}}.
\eex
Notice that $\arcsin\sbkt{(1-\rho)/(1+\rho)}$ is a decreasing function of $\rho$ on the interval $ \interval[open]{0}{1}$. This indicates a physical interpretation that a coarse quantizer introduces a large phase. As an example, setting $\rho={1}/{3}$ gives the corresponding nonlinearity sector from ${1/2}$ to ${{3}/{2}}$ and the phase sector $\Phi(\sysn)\subseteq \interval{-\pi/6}{\pi/6}$.
\end{example}

\vspace{-6mm}

\subsubsection{From Passivity Indices to a Phase Sector}
We provide the phase estimation for a more general class of nonlinear systems. Recall that a very strictly passive system $\sysp\colon\ltep \rightarrow \ltep$ satisfies
\be\label{eq:VSP_copy}
\ininf{u}{\sysp u}\geq \delta \norm{u}^2_2+\epsilon \norm{\sysp u}^2_2,\quad \forall u \in \ltp,
\ee
where $\delta>0$ and $\epsilon>0$ are called the input passivity index and output passivity index, respectively. Notice that, the sector bounded static nonlinearity satisfying (\ref{eq:sector}), according to (\ref{eq:Lem1}) in the proof of Proposition~\ref{lem:sector} in Appendix, is a special very strictly passive system, with the indices $\delta={ab}/{(a+b)}$ and $\epsilon={1}/{(a+b)}$. Considering that the phase of the sector bounded static nonlinearity can be nicely estimated using the nonlinearity sector, we ask the following question: Can we estimate the phase of the very strictly passive system using the two indices? The answer is yes, as detailed in the next proposition.
\begin{proposition}\label{lem:VSP}
For a very strictly passive system $\sysp$ in (\ref{eq:VSP_copy}),
 the phase sector of $\sysp$ satisfies
\bex
 {\Phi}(\boldsymbol{P})\subseteq\interval{-{\arcsin\sqrt{1-4\delta\epsilon}}}{\arcsin\sqrt{1-4\delta\epsilon}}.
\eex
\end{proposition}

Clearly, Proposition~\ref{lem:sector} can be recovered by taking the two indices as $\delta={ab}/{(a+b)}$ and $\epsilon={1}/{(a+b)}$ in Proposition~\ref{lem:VSP}. It is also noteworthy that the two indices $\delta$ and $\epsilon$ in Proposition~\ref{lem:VSP} are always in the domain of ${\arcsin\sqrt{1-4\delta\epsilon}}$, since, for a very strictly passive system,  $\delta$ and $\epsilon$ always satisfy the constraint $\delta\epsilon\in \interval[open left]{0}{1/4}$, as stated in \cite[Lemma~2.6]{Yu:13}.

\subsection{The Phase of MIMO LTI Systems}
When $\sysp$ is restricted to be a MIMO LTI system with transfer function ${P}(s)\in \rhinf^{n\times n}$, our time-domain phase definition reduces to the frequency-domain definition in \cite{Chen:19, Chen:21}. To clarify this connection, we first present the following proposition which reveals that the angular numerical range defined in the time domain (\ref{eq:nrange}) has an equivalent frequency-domain representation. This representation is utilized in \cite{Chen:19, Chen:21} to define the MIMO LTI system phase.

\begin{proposition}\label{prop:LTI}
For a causal stable LTI system $\sysp$ with transfer function ${P}(s) \in \rhinf^{n\times n}$, it holds that
\be\label{eq:Prop_pf3}
 \text{cl}\bbkt{W^{\prime}(\sysp)} = \text{cl}~\text{conv}\bbkt{x^* {P}(\jw) x \in \mathbb{C} \mid \omega \in\interval{0}{\infty}, x \in \cn}
 \ee
 where $\text{cl}\bbkt{\cdot}$ and $\text{conv}\bbkt{\cdot}$ denote the closure and convex hull of a set, respectively.
\end{proposition}

In (\ref{eq:Prop_pf3}), it is noteworthy that only the frequency response over positive frequencies is involved. The fundamental reason lies in the fact that the frequency response of an LTI system $P(s) \in \rhinf^{n\times n}$ is conjugate symmetric, namely,  $P(\jw)=\overline{P(-\jw)}$. Consequently, its positive spectrum encodes all available information and its negative spectrum is ``redundant'' as a mathematical artifact. This ``redundant'' component is handily discarded by utilizing the analytic signal, as shown in the proof of Proposition~\ref{prop:LTI} in Appendix. Notice that the system properties over positive frequencies are also highlighted in the definitions of LTI counterclockwise systems \cite{Angeli:06}, negative imaginary systems \cite{Petersen:10} and sectorial systems \cite{Chen:19, Chen:21}.

Next, we provide an equivalent characterization for MIMO LTI semi-sectorial systems accordingly based on the positive spectrum in Proposition~\ref{prop:LTI}. For a causal stable LTI system $\sysp$ with $P(s)\in \rhinf^{n\times n}$, we denote the set
\bex
W^{\prime}(P(s))\coloneqq \text{cl}~\text{conv}\bbkt{x^* {P}(\jw) x \in \mathbb{C} \mid \omega \in\interval{0}{\infty}, x \in \cn}.
\eex
By Proposition~\ref{prop:LTI}, it holds that $\text{cl}\bbkt{W^{\prime}(\sysp)}=W^{\prime}(P(s))$. Therefore, $P(s)$ is semi-sectorial if and only if $W^{\prime}(P(s))$ is contained in a closed complex half-plane. In addition, it is sectorial if and only if further (\ref{eq:SS}) holds. Subsequently, the notions $\underline{\phi}(P(s)), \overline{\phi}(P(s))$ and $\Phi(P(s))$ for $P(s)$ can be obtained accordingly.  In parallel, the more general multiplier-based version for $P(s)$, i.e.,
{$\mulpi$-semi-sectorial} ($\mulpi$-sectorial) LTI systems with {$\mulpi$-phases} $\Phi_{\mulpi}(P(s))$, can be characterized similarly to the above. The following proposition summarizes frequency-domain features of the aforementioned LTI systems.

\begin{proposition}\label{prop:LTI_sectorial}
  For a causal stable LTI system $\sysp$ with $P(s)\in \rhinf^{n\times n}$ and a bounded LTI multiplier $\mulpi$ with $\Pi(s)\in \linf^{n\times n}$, the following statements are true:
  \begin{enumerate}
  \item [(a)\label{item:01}] $P(s)$ is sectorial (semi-sectorial, resp.) if and only if there exist $\alpha \in \interval{-\pi}{\pi}$ and $\epsilon>0$ such that \bex
\text{He}\rbkt{e^{j\alpha}P(\jw)}\geq 2\epsilon P(\jw)^*P(\jw)~(\geq 0,~\text{resp.})
\eex
holds for all $\omega\in\interval{0}{\infty}$.
   \item [(b)] The set $\text{cl}\bbkt{W_{\mulpi}^{\prime}(\sysp)}$ is equal to the set
\bex
  \text{cl}~\text{conv}\bbkt{x^* \Pi(\jw)^*{P}(\jw) x \in \mathbb{C} \mid \omega\in\interval{0}{\infty}, x \in \cn}.\eex
    \item [(c)] $P(s)$ is $\mulpi$-sectorial ($\mulpi$-semi-sectorial, resp.) if and only if there exist $\alpha \in \interval{-\pi}{\pi}$ and $\epsilon>0$ such that \bex
\text{He}\rbkt{e^{j\alpha}\Pi(\jw)^*P(\jw)}\geq 2\epsilon P(\jw)^*P(\jw)~(\geq 0,~\text{resp.})\eex
holds for all $\omega\in\interval{0}{\infty}$.
  \end{enumerate}
\end{proposition}

Statement (a) of Proposition~\ref{prop:LTI_sectorial} shows that the semi-sectorial and sectorial LTI systems generalize the sectorial system defined in \cite{Chen:19, Chen:21}. In what follows, we further demonstrate that the notion of frequency-wise sectorial systems proposed in \cite{Chen:19, Chen:21} can be covered by the $\mulpi$-sectorial LTI systems. Concretely, in \cite{Chen:19, Chen:21}, $P(s)\in \rhinf^{n\times n}$ is called \textit{frequency-wise sectorial}
if, for each $\omega\in \interval{-\infty}{\infty}$, the frequency-dependent set \bex\bbkt{x^*P(\jw)x  \in \mathbb{C} \mid x \in \cn~\text{with}~\abs{x}=1},\eex
namely, the matrix numerical range of $P(\jw)$, is contained in an open complex half-plane. Then, for each $\omega$, the frequency-wise phase notions $\underline{\phi}(P(\jw)), \overline{\phi}(P(\jw))$ and $\Phi(P(\jw))$ are defined on the basis of the matrix numerical range of $P(\jw)$. Here, to avoid cluttering the notation for LTI systems, we distinguish the frequency-wise phase $\Phi(P(\jw))$ from $\Phi(P(s))$. Notice that the frequency-wise version is less conservative than the previous version in statement (a) of Proposition~\ref{prop:LTI_sectorial}. In the former, the set may be contained in different half-planes for different frequencies, while in the latter, a uniform half-plane is required for all positive frequencies. This deficiency can be made up for using appropriate multipliers. The following proposition indicates that a frequency-wise sectorial system is a special $\mulpi$-sectorial system.
\begin{proposition}\label{prop:freq_sectorial}
For a frequency-wise sectorial system $P(s)\in \rhinf^{n\times n}$, there exists a multiplier $\mulpi$ with $\Pi(s)\in \linf$ such that $P(s)$ is $\mulpi$-sectorial.
\end{proposition}

We can also deduce, in the same manner as before, that the \textit{frequency-wise semi-sectorial} system in \cite{Chen:19, Chen:21} is a special $\mulpi$-semi-sectorial system.

\section{Nonlinear Small Phase Theorems}\label{sec:04}
\begin{figure}[htb]
\centering
\setlength{\unitlength}{0.8mm}
\begin{picture}(50,25)
\thicklines \put(0,20){\vector(1,0){8}} \put(10,20){\circle{4}}
\put(12,20){\vector(1,0){8}} \put(20,15){\framebox(10,10){$\boldsymbol{P}$}}
\put(30,20){\line(1,0){10}} \put(40,20){\vector(0,-1){13}}
\put(38,5){\vector(-1,0){8}} \put(40,5){\circle{4}}
\put(50,5){\vector(-1,0){8}} \put(20,0){\framebox(10,10){$\boldsymbol{C}$}}
\put(20,5){\line(-1,0){10}} \put(10,5){\vector(0,1){13}}
\put(5,10){\makebox(5,5){$y_2$}} \put(40,10){\makebox(5,5){$y_1$}}
\put(0,20){\makebox(5,5){$e_1$}} \put(45,0){\makebox(5,5){$e_2$}}
\put(13,20){\makebox(5,5){$u_1$}} \put(32,0){\makebox(5,5){$u_2$}}
\put(10,10){\makebox(6,10){$-$}}
\end{picture}\caption{A standard feedback system $\gof$.} \label{fig:feedback}
\end{figure}
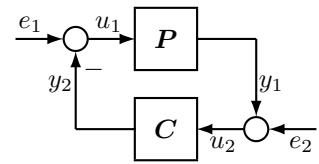

Having defined the nonlinear system phase, we proceed to stability analysis of feedback interconnected semi-sectorial and/or sectorial systems. Consider the standard feedback system shown in Fig.~\ref{fig:feedback}, where $\sysp \colon \ltep \rightarrow \ltep$ and $\sysc \colon \ltep \rightarrow \ltep$ are two causal stable systems, $e_1$ and $e_2$ are external signals, and $u_1, u_2, y_1$ and $y_2$ are internal signals. Let $\gof$ denote the feedback system. Algebraically, we have the following equations:
\be\label{eq:feedback}
u=e-\tbt{0}{\boldsymbol{I}}{-\boldsymbol{I}}{0}y\quad\text{and}\quad y=\tbt{\boldsymbol{P}}{0}{0}{\boldsymbol{C}}u,
\ee
where $u=\left[u_1~u_2\right]^T$, $e=\left[{e_1}~{e_2}\right]^T$, $y=\left[{y_1}~{y_2}\right]^T$ and $\sysi$ denotes the identity system.

 We introduce two indispensable definitions on feedback systems. Firstly, the well-posedness of $\gof$ is an important assumption to guarantee that the closed-loop system (\ref{eq:feedback}) makes sense as a model of a real system. We stipulate the following definition from \cite[Section~4.2]{Willems:71}, and assume all the feedback systems in this paper are well-posed.
\begin{definition}\label{def:wellposed}
   The feedback system $\gof$ is said to be well-posed if $
       u \mapsto e \colon \ltep \rightarrow \ltep \coloneqq \tbt{\sysi}{\sysc}{-\sysp}{\sysi}$ has a causal inverse on $\ltep$.
\end{definition}
Secondly, for a well-posed feedback system, we are interested in the following feedback stability.
\begin{definition}
    A well-posed feedback system $\gof$ is said to be stable if there exists a constant $c>0$ such that, $
    \norm{\boldsymbol{\Gamma}_T u}_2 \leq c \norm{\boldsymbol{\Gamma}_T e}_2$
    for all $T\geq 0$ and for all $e\in \ltep$.
\end{definition}

Given the aforementioned definitions, recall a fundamental version of the classical nonlinear small gain theorem \cite{Zames:66}, \cite[Section~2.1]{Van:17}: For causal stable systems $\sysp$ and $\sysc$, the well-posed feedback system $\gof$ is stable if
\be \label{eq:small-gain}
\norm{\sysp}\norm{\sysc}<1.
\ee
The elegant inequality (\ref{eq:small-gain}) is known as the small gain condition.  Keeping this in mind, we are ready to present our main result, a parallel stability condition given by the nonlinear system phases.

\begin{theorem}[Nonlinear small phase theorem]\label{Thm:nsp}
  For sectorial $\sysp$ and semi-sectorial $\sysc$, the well-posed feedback system $\gof$ is stable if
\begin{align}\label{eq:small-phase}
\begin{aligned}
  {\overline{\phi}}(\sysp) + {\overline{\phi}}(\sysc) &< \pi,\\
{\underline{\phi}}(\sysp) + {\underline{\phi}}(\sysc) &> -\pi.
\end{aligned}
\end{align}
\end{theorem}

Theorem~\ref{Thm:nsp} also holds if $\boldsymbol{C}$ is sectorial and $\boldsymbol{P}$ is semi-sectorial. Moreover, Theorem~\ref{Thm:nsp} still holds under the stronger stability in the following sense:  For sectorial $\sysp$ and semi-sectorial $\sysc$, the well-posed feedback system  $\rbkt{\tau\boldsymbol{P}}\,\#\,\boldsymbol{C}$ is stable for all $\tau>0$ if (\ref{eq:small-phase}) holds. A one-line proof follows from that $\Phi(\tau\sysp)=\Phi(\sysp)$ for all $\tau>0$. This stronger stability coincides with the concept of infinite gain margin in classical control theory.

Theorem~\ref{Thm:nsp} provides a crucial condition (\ref{eq:small-phase}), called the small phase condition, from a phasic viewpoint to guarantee the closed-loop stability.  Recall that the small gain condition (\ref{eq:small-gain}) requires that the loop $\lt$-gain $\norm{\sysp}\norm{\sysc}$ be less than one. By contrast, the small phase condition (\ref{eq:small-phase}) requires that, the loop phase supremum $\overline{\phi}(\boldsymbol{P})+\overline{\phi}(\boldsymbol{C})$ be less than $\pi$, while the loop phase infimum $\underline{\phi}(\boldsymbol{P})+\underline{\phi}(\boldsymbol{C})$ be greater than $-\pi$. Consequently, the surplus or shortage of $\Phi(\sysp)$, in comparison to $\pi$ or $-\pi$, can be compensated by $\Phi(\sysc)$, as long as (\ref{eq:small-phase}) holds. Very often a physical system, which we need to deal with in practice, is purely phase-lag, for example, a system $\sysp$ with $\Phi(\boldsymbol{P})=\interval{-4\pi/3}{-2\pi/3}$. In addition, its phase infimum ${\underline{\phi}}(\boldsymbol{P})$ is short of $-\pi$. To handle with such a $\sysp$, one can conveniently adopt lead compensation, e.g., using a lead compensator $\sysc$ with $
  \Phi(\boldsymbol{C})=\interval{2\pi/5}{4\pi/5}$, to drag the loop phase infimum back to be greater than $-\pi$. This idea coincides with phase lead-lag compensation techniques in classical control theory.

Introducing multipliers can reduce the conservatism of Theorem~\ref{Thm:nsp}. Its motivation is twofold. First, for those systems that are not semi-sectorial and thus we cannot define their phases, fortunately, it is possible to find suitable multipliers $\mulpi$ to make them $\mulpi$-semi-sectorial, and then define the corresponding $\mulpi$-phases. Second, when dealing with MIMO LTI systems, the use of multipliers in this study can recover the frequency-wise analysis in the literature \cite{Chen:19, Chen:21}, as partially elaborated in Proposition~\ref{prop:freq_sectorial}. Therefore, we present the following generalized version of Theorem~\ref{Thm:nsp} involving multipliers.

\begin{theorem}[Generalized nonlinear small phase theorem]\label{thm:gspt}
   For causal stable systems $\sysp$ and $\sysc$, the well-posed feedback system $\gof$ is stable if there exists a bounded LTI multiplier $\mulpi$ with transfer function $\Pi(s) \in \linf^{n\times n}$ such that,
$\sysp$ is $\mulpi$-sectorial, $\sysc$ is $\mulpi^*$-semi-sectorial and
\begin{align*}\label{eq:G_small-phase}
\begin{aligned}
  {\overline{\phi}_{\mulpi}}(\sysp) + {\overline{\phi}_{\mulpi^*}}(\sysc) &< \pi,\\
{\underline{\phi}_{\mulpi}}(\sysp) + {\underline{\phi}_{\mulpi^*}}(\sysc) &> -\pi.
\end{aligned}
\end{align*}
\end{theorem}

Some existing results in the literature can be subsumed into the nonlinear small phase theorems. First, when $\boldsymbol{P}$ is output strictly passive and $\boldsymbol{C}$ is stable passive, Theorem~\ref{Thm:nsp} reduces to a version of the passivity theorem \cite[Section~6.6.2]{Vidyasagar:93}. Second, when dealing with MIMO LTI systems in $\rhinf^{n\times n}$, Theorem~\ref{Thm:nsp} generalizes the $\hinf$-phase version LTI small phase theorem in \cite{Chen:19, Chen:21} where sectorial systems are concerned. Meanwhile, Theorem~\ref{thm:gspt} further extends the frequency-wise version in \cite{Chen:19, Chen:21} where frequency-wise sectorial and semi-sectorial systems are involved. Specifically, we next state two corollaries for MIMO LTI systems, which are derived from Theorems~\ref{Thm:nsp} and \ref{thm:gspt}, respectively. The first one is as follows.

\begin{corollary}[LTI small phase theorem]\label{coro:LTI}
  For sectorial $P(s)\in \rhinf^{n\times n}$ and semi-sectorial $C(s)\in \rhinf^{n\times n}$, the well-posed feedback system ${P}\,\#\,{C}$ is stable if
\begin{align*}
  {\overline{\phi}}(P(s)) + {\overline{\phi}}(C(s)) &< \pi,\\
  {\underline{\phi}}(P(s)) + {\underline{\phi}}(C(s))&> -\pi.
\end{align*}
\end{corollary}

Corollary~{\ref{coro:LTI}} is conservative in the sense that it does not make full use of the characteristic frequency-wise analysis of LTI systems. As mentioned before, one of the advantages of Theorem~\ref{thm:gspt} is that,  by choosing a suitable multiplier $\mulpi$, we can derive a frequency-wise result, as detailed in the next corollary.
\begin{corollary}[Frequency-wise LTI small phase theorem]\label{coro:freq_LTI}
  For frequency-wise sectorial $P(s)\in \rhinf^{n\times n}$ and frequency-wise  semi-sectorial $C(s)\in \rhinf^{n\times n}$, the well-posed feedback system ${P}\,\#\,{C}$ is stable if, for all $\omega\in \interval{-\infty}{\infty}$,
\begin{align*}
  {\overline{\phi}}(P(\jw)) + {\overline{\phi}}(C(\jw)) &< \pi,\\ {\underline{\phi}}(P(\jw)) + {\underline{\phi}}(C(\jw))&> -\pi.
\end{align*}
\end{corollary}

Corollary~\ref{coro:freq_LTI} has been proved in \cite{Chen:21}. We provide an alternative proof in Appendix based on Theorem~\ref{thm:gspt}.

\subsection{An Application to the Lur'e Systems}\label{sec:04:a}
The well-known Lur'e problem \cite[Section~5.6]{Vidyasagar:93} concerns the feedback interconnection of an LTI system and a sector bounded static nonlinearity. It aims at deriving a closed-loop stability condition against all static nonlinearities bounded in a sector. Over the past half century, the celebrated circle criterion \cite{Sandberg:64,Zames:66_2} and Popov criterion \cite[Section~6.6]{Vidyasagar:93} have stood out. An appealing aspect of these criteria is their geometric interpretations when scalar systems are considered. In Proposition~\ref{lem:sector}, we have shown that the phase of a sector bounded static nonlinearity is well estimated. Consequently, we can apply the nonlinear small phase theorem to the Lur'e system, which shares a similar flavor to the geometric interpretation of the circle criterion. This similarity will become clear as we proceed.

Consider a simple Lur'e system $\sysp\,\#\,{\sysn}$ which consists of a SISO LTI system $\sysp$ with $P(s)\in \rhinf$ and a scalar static system $\boldsymbol{N}$ satisfying the nonlinearity sector condition (\ref{eq:sector}). The following input-output version of the circle criterion is adopted from \cite[Section~6.6.1]{Vidyasagar:93}.

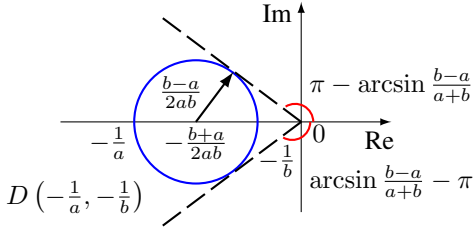
\begin{figure}[htb]
\centering
\setlength{\unitlength}{0.8mm}
\begin{picture}(60,40)
\put(0,20){\vector(1,0){55}}
\put(40,5){\vector(0,1){35}}
\thicklines
\put(43,18){\makebox(0,0){$0$}}
\put(0,5){\makebox(5,5){$D\rbkt{-\frac{1}{a}, -\frac{1}{b}}$}}
{\color{blue}\put(22.5,20){\circle{20.5}}}
\put(8,16.5){\makebox(0,0){$-\frac{1}{a}$}}
\put(36,13.5){\makebox(0,0){$-\frac{1}{b}$}}
\put(22.5,16.5){\makebox(0,0){$-\frac{b+a}{2ab}$}}
\put(20,25){\makebox(0,0){$\frac{b-a}{2ab}$}}
\put(55,26){\makebox(0,0){$\pi-\arcsin\frac{b-a}{a+b}$}}
\put(55,10){\makebox(0,0){$\arcsin\frac{b-a}{a+b}-\pi$}}
\multiput(40,20)(-4,3){6}{\line(-4,3){3}}
\multiput(40,20)(-4,-3){6}{\line(-4,-3){3}}
\thicklines
\put(22.5,20){\vector(3,4){6.3}}
{\color{red}\put(39, 20){\arc[0,122]{3}}
\put(38.5, 20){\arc[-120,0]{3}}}
\put(53,17){\makebox(0,0){$\rep$}}
\put(36,38){\makebox(0,0){$\imp$}}
\end{picture}\caption{The interpretations of the circle criterion with the disk $D(-1/a, -1/b)$ (blue) and the nonlinear small phase theorem with two rays (dashed) for the Lur'e system when $b>a>0$.}\label{fig:circle}
\end{figure}

\begin{lemma}[Circle criterion]\label{lem:lure}
The well-posed Lur'e system ${\sysp}\,\#\,{\sysn}$ is stable if the Nyquist plot of $P(s)$ is bounded away from the disk
 \bex
  D \rbkt{-\frac{1}{a}, -\frac{1}{b}} =  \bbkt{z\in \mathbb{C}\Bigg| ~\abs{z + \frac{a+b}{2ab}} \leq \frac{b-a}{2ab}},\eex
   namely,
  \bex
  \inf_{\substack{\omega\in\interval{-\infty}{\infty},\\z\in D\rbkt{-\frac{1}{a}, -\frac{1}{b}}}} \abs{P(\jw)-z}>0.
  \eex
\end{lemma}

As illustrated by Fig.~\ref{fig:circle}, the circle criterion indicates that the blue disk $D(-{1}/{a}, -{1}/{b})$ is the forbidden region for the Nyquist plot of $P(s)$. It is noteworthy that the disk $D(a, b)$ in Proposition~\ref{lem:sector} is exactly $-ab$ times the disk $D(-{1}/{a}, -{1}/{b})$ in Lemma~\ref{lem:lure}. In addition, $\max \angle D(a,b)$ and $\max \angle D(-{1}/{a}, -{1}/{b})$ are a pair of supplementary angles, i.e., \bex \max \angle D(a,b) + \max \angle D(-{1}/{a}, -{1}/{b})=\pi.\eex
In contrast to the circle criterion, we can obtain a new forbidden region by virtue of the nonlinear small phase theorem and Proposition~\ref{lem:sector}. To be specific, we have the following corollary.
\begin{corollary}\label{coro:lure}
 The well-posed Lur'e system $\rbkt{\tau {\sysp}}\,\#\,{\sysn}$ is stable for all $\tau >0$ if $\sysp$ is semi-sectorial and
 \be\label{eq:coro2}
\angle P(\jw) \in  \interval[open]{\arcsin\frac{b-a}{b+a}-\pi}{\pi-\arcsin\frac{b-a}{b+a}}
 \ee
 for all $\omega\in\interval{-\infty}{\infty}$ and $P(\jw)\neq 0$.
\end{corollary}

In Corollary~\ref{coro:lure}, the phasic condition (\ref{eq:coro2}) offers a new forbidden region.  As depicted in Fig.~\ref{fig:circle}, this region is a convex cone spanned by the disk $D(-{1}/{a}, -{1}/{b})$, which opens to the left and is characterized by two dashed rays. Consequently, the condition (\ref{eq:coro2}) can be restated in the sense that the Nyquist plot of $P(s)$ is bounded away from the new region.

In Fig.~\ref{fig:circle}, observe that the new region (dashed) is more conservative than the region (blue) in the circle criterion, but they share the same angles (red) of the boundaries of regions. The underlying reason is that the former is determined by only phasic information, while the latter is not. In other words, gain information is mixed in the circle criterion. For a fair comparison,  we rule out this information by imposing a stronger stability requirement on ${\sysp}\,\#\,{\sysn}$, namely,  $\rbkt{\tau {\sysp}}\,\#\,{\sysn}$ should be stable for all $\tau>0$. Roughly speaking, ``infinite gain margin'' is required. Then, the forbidden region in the circle criterion will be enlarged and become a cone formed by infinitely many disks. Under such a circumstance, it is worth noting that for a semi-sectorial $\sysp$, Corollary~\ref{coro:lure} is equivalent to the circle criterion.

\section{The Phases of Interconnected Systems}\label{sec:Parallel}
A large-scale nonlinear network is often composed of a large number of subsystems. When these subsystems are passive, there is a beneficial property that either a parallel or negative feedback interconnected system is still passive. This property contributes to scalable analysis and synthesis in the large-scale nonlinear network. Inspired by this, in what follows, we study the phases of the parallel and feedback interconnections, respectively.

We start from a parallel interconnection $\sysp$ of two given systems $\sysp_1$ and $\sysp_2$, namely, $\sysp=\sysp_1+\sysp_2$.
Let two real numbers $\alpha, \beta\in \interval{-3\pi/2}{3\pi/2}$, where $\beta-\alpha \in \interval[open]{0}{\pi}$. We consider the following set of phase-bounded systems
\begin{multline*}
 \mathcal{P}(\alpha,\beta)\coloneqq   \left\{ \sysp \colon \ltep\to\ltep \big| \right.\\
 \left. \sysp~\text{is sectorial and}~\Phi(\sysp)\subseteq \interval{\alpha}{\beta} \right\}.
\end{multline*}
The next proposition reveals a nice property of the set $\mathcal{P}(\alpha,\beta)$.
\begin{proposition}\label{prop:parallel}
The set $\mathcal{P}(\alpha,\beta)$ is a convex cone.
\end{proposition}

Proposition~\ref{prop:parallel} implies that, for two systems $\sysp_1\in \mathcal{P}(\alpha,\beta)$ and $\sysp_2\in \mathcal{P}(\alpha,\beta)$, the parallel interconnection $\sysp=\sysp_1+\sysp_2$ satisfies $\sysp\in \mathcal{P}(\alpha,\beta)$. Proposition~\ref{prop:parallel} also holds for phase-bounded semi-sectorial systems, and for phase-bounded  $\mulpi$-semi-sectorial ($\mulpi$-sectorial) systems.

Next, we investigate how the phases of closed-loop systems are related by those of open-loop systems. For a stable feedback system $\gof$ in Fig.~\ref{fig:feedback}, we assume $e_2=0$, and then let $\boldsymbol{G}_1$ denote the closed-loop map described by $
\boldsymbol{G}_1\coloneqq e_1 \mapsto y_1 \colon \ltep \rightarrow \ltep.$ Likewise, we assume $e_1=0$, and then denote $\boldsymbol{G}_2\coloneqq e_2 \mapsto y_2 \colon \ltep \rightarrow \ltep$. The following proposition indicates that $\Phi(\boldsymbol{G}_1)$ and $ \Phi(\boldsymbol{G}_2)$ can be well estimated from $\Phi(\sysp)$ and $\Phi(\sysc)$.
\begin{proposition}\label{prop:CL_phase}
For a stable feedback system $\gof$ with semi-sectorial systems $\sysp$ and $\sysc$, we assume that
\begin{align*}
  {\overline{\phi}}(\sysp) + {\overline{\phi}}(\sysc) \leq \pi\quad \text{and}\quad {\underline{\phi}}(\sysp) + {\underline{\phi}}(\sysc) \geq -\pi.
\end{align*}
Then, we have
\begin{align*}
  \Phi(\boldsymbol{G}_1) &\subseteq \interval {\min\bbkt{\underline{\phi}(\sysp),-\overline{\phi}(\sysc)}}{\max\bbkt{\overline{\phi}(\sysp),-\underline{\phi}(\sysc)}},\\
   \Phi(\boldsymbol{G}_2) &\subseteq \interval {\min\bbkt{\underline{\phi}(\sysc),-\overline{\phi}(\sysp
   )}}{\max\bbkt{\overline{\phi}(\sysc),-\underline{\phi}(\sysp)}}.
\end{align*}
\end{proposition}

Proposition~\ref{prop:CL_phase} also holds when $\sysp$ is $\mulpi$-semi-sectorial and $\sysc$ is $\mulpi^*$-semi-sectorial. Propositions~\ref{prop:parallel}~and~\ref{prop:CL_phase} generalize the results of the parallel and feedback interconnections of stable passive systems.

By virtue of Propositions~\ref{prop:parallel}~and~\ref{prop:CL_phase}, if a large-scale network consists of subsystems via the appropriate parallel and negative feedback interconnections, then its phase can be simply obtained by the phases of subsystems. This manifests the advantage of scalability, and offers us a starting point to study a large-scale network using the nonlinear system phase theory.

\section{Connections of the Nonlinear System Phase to Existing Notions}\label{sec:05}
In this section, we attempt to connect the nonlinear system phase and small phase theorem to three crucial notions in the literature, namely, the dissipativity \cite{Willems:72, Hill:80}, integral quadratic constraints (IQCs) \cite{Megretski:97} and multipliers\cite{Zames:68, Desoer:75}.  We also reveal a physical interpretation of the phase concerning the real and reactive energy/power. To avoid duplication of the specifications of a nonlinear system $\sysp$ in the rest of this section, we stipulate that $\sysp\coloneqq u\mapsto y\colon \ltep \to \ltep$ is causal and stable.

\subsection{The Dissipativity and IQCs}\label{sec:dissipativity}
It is well known that the passivity and $\lt$-gain can be incorporated in a unified dissipativity framework. Can the nonlinear system phase be incorporated so?  The answer will be made clear as we proceed.  The property of dissipativeness can be regarded as a state-space \cite{Willems:72} or input-output property \cite{Hill:80}. The latter is adopted in this paper. The notion of supply rates, as an abstraction of the concept of physical input power, plays a key role in the input-output dissipativity theory.  We adopt the commonly-used $\rbkt{Q, S, R}$-supply rate from \cite{Hill:80}, namely, the function $s\colon \rn \times \rn \to \mathbb{R}$ given by
\be\label{eq:def_srate}
 s(u(t), y(t)) \coloneqq \tbo{u(t)}{y(t)}^T \tbt{R}{S}{S^T}{Q} \tbo{u(t)}{y(t)},
\ee
 where $Q, S, R \in \mathbb{R}^{n\times n}$ are constant matrices, with $Q=Q^T$ and $R=R^T$ symmetric. The supply rate $s(u(t), y(t))$ is evaluated along the system input and output at time $t$ as a quadratic form. Then, a system $\boldsymbol{P}$ is ultimately dissipative \cite[Definition~3]{Hill:80} with respect to the $\rbkt{Q, S, R}$-supply rate $s({u}(t), {y}(t))$ if and only if
 \bex
  \int_{0}^{\infty} s({u}(t), {y}(t)) dt \geq 0,\quad \forall u\in \ltp.
 \eex
This integral of the supply rate is a measure of energy. Roughly speaking, an ultimately dissipative system dissipates energy from the initial time zero up to the final time infinity. The energy here may correspond to real physical energy, but in most cases, it is a mathematical abstraction of energy.

The passive or gain-bounded systems can be characterized as ultimately dissipative systems using the $\rbkt{Q, S, R}$-supply rates. Firstly, a system $\boldsymbol{P}$ is passive if it is ultimately dissipative with respect to the $\rbkt{0, I/2, 0}$-supply rate,  where $I$ is the identity matrix. For some real systems, this supply rate corresponds to usual physical power. For example, it represents electric power in resistor-inductor-capacitor circuits when the inputs and outputs are taken to be voltages and currents, respectively. Secondly, a system $\boldsymbol{P}$ has its $\lt$-gain no greater than $\gamma$ if it is ultimately dissipative with respect to the $\rbkt{-I/2, 0, \gamma^2 I/2}$-supply rate, where $\gamma>0$.

In an attempt to relate the phase with the dissipativity, we first present the following  proposition, which establishes an equivalence between the semi-sectorial systems and corresponding energy inequality.
\begin{proposition}\label{prop:TDI}
A system $\sysp$ is semi-sectorial  if and only if there exists a constant $\alpha \in \interval{-{\pi}}{{\pi}}$ such that
 \be\label{eq:TDI}
 \ininf{u}{\cos\alpha y - \sin\alpha \boldsymbol{H}y}\geq 0,\quad \forall u \in \ltp.
 \ee
\end{proposition}
To make (\ref{eq:TDI}) more compact,  let
\be\label{eq:auxiliary_M}
\boldsymbol{M}\colon \lt\to\lt \coloneqq {\cos\alpha \sysi - \sin\alpha \sysh}
\ee
and rewrite (\ref{eq:TDI}) in the following form: $\ininf{u}{\sysm y}\geq 0$ for all $u\in\ltp$.
 Clearly, this crucial $\sysm$ is a noncausal and bounded LTI operator. Notice that $Q, S$ and $R$ in the supply rate (\ref{eq:def_srate}) are constant matrices, i.e., memoryless LTI operators. Recall the aforementioned question: Can the nonlinear system phase be characterized by a certain supply rate? The answer is affirmative if $Q, S$ and $R$ can be extended to be dynamic LTI operators. Specifically, let $\boldsymbol{Q}, \boldsymbol{S}, \boldsymbol{R} \colon \lt \to \lt$  be bounded (possibly noncausal) LTI operators with transfer functions $Q(s), S(s), R(s)\in \linf$, respectively, with $\boldsymbol{Q}=\boldsymbol{Q}^*$ and $\boldsymbol{R}=\boldsymbol{R}^*$ self-adjoint. This natural extension yields the so-called \textit{dynamic} $(\boldsymbol{Q}, \boldsymbol{S}, \boldsymbol{R})$-\textit{supply rate} and corresponding \textit{ultimate} $(\boldsymbol{Q}, \boldsymbol{S}, \boldsymbol{R})$-\textit{dissipativity}. Let $\alpha\in \interval{-\pi}{\pi}$. With these preparations, we then unveil the dynamic supply rate describing the phase-bounded systems, as elaborated below.

\begin{proposition}\label{co:2}
A system $\boldsymbol{P}$ has its phase $\Phi(\sysp)$ contained in $\interval{-{\pi}/{2}-\alpha}{{\pi}/{2}-\alpha}$ if and only if it is ultimately dissipative with respect to the dynamic $\rbkt{0, \sysm/2, 0}$-supply rate, i.e.,
\be\label{eq:supply}
   s(u(t), y(t))=u(t)^T \sbkt{\cos\alpha y(t) - \sin\alpha (\boldsymbol{H}y)(t)}.
\ee
\end{proposition}

In Proposition~\ref{co:2}, the phase of a system belongs to a spread-$\pi$ sector $\interval{-{\pi}/{2}-\alpha}{{\pi}/{2}-\alpha}$. When the more accurate phase information is available, a generalization can be made by intersecting a few spread-$\pi$ phase sectors. For example, a system has its phase contained in $\interval{-\pi/4}{\pi/3}$ if and only if, it is ultimately dissipative with respect to the two dynamic supply rates, namely, $\alpha=\pi/6$ and $\alpha=-\pi/4$ in (\ref{eq:supply}).

The kind of dynamic supply rate (\ref{eq:supply}) is new, while the notion of dynamic supply rates is not. It is pointed out in \cite{Willems:98, Willems:02} that there are many examples where the supply rate is given not by a static function of the external variables, but by a quadratic differential form of them. Other types of dynamic supply rates, such as the counterclockwise dynamics \cite{Angeli:06} from an input-output perspective, differential passivity \cite{Forni:13, Van:13} and quadratic dynamic supply rate \cite[Chapter 8]{Arcak:16} from a state-space perspective, are also examined in the literature.

A closely related notion of the dynamic ultimate $(\boldsymbol{Q}, \boldsymbol{S}, \boldsymbol{R})$-dissipativity is the IQC \cite{Megretski:97}. The IQC theory is also a broad framework that unifies the passivity and $\lt$-gain, within which noncausal dynamic multipliers may be accommodated. Concretely, a system $\sysp$ is said to satisfy the IQC defined by a self-adjoint LTI operator $\boldsymbol{\Psi}=\boldsymbol{\Psi}^*$ with $\Psi(s)\in\linf^{2n\times 2n}$, a.k.a. a multiplier, if
\bex
\int_{-\infty}^{\infty}{\tbo{\hat{u}(\jw)}{\hat{y}(\jw)}}^*{{\Psi(\jw)}\tbo{\hat{u}(\jw)}{\hat{y}(\jw)}} d\omega\geq 0,\quad\forall u\in\ltp.
\eex
Then, the dynamic ultimate  $(\boldsymbol{Q}, \boldsymbol{S}, \boldsymbol{R})$-dissipativity is equivalent to the IQC defined by the multiplier $\boldsymbol{\Psi}$ via $\boldsymbol{\Psi}= \tbt{\boldsymbol{R}}{\boldsymbol{S}}{\boldsymbol{S}^*}{\boldsymbol{Q}}$. We refer the interested reader to \cite{Seiler:15, Scherer:18, Khong:21} for the close relationship between the IQC and dissipativity.

How to find meaningful multipliers $\boldsymbol{\Psi}$ is a key question in the IQC theory. It is known that the passivity and $\lt$-gain are connected with the two static multipliers $\boldsymbol{\Psi}=\tbt{0}{\boldsymbol{I}}{\boldsymbol{I}}{0}$ and  $\boldsymbol{\Psi}=\tbt{\gamma^2\sysi}{0}{0}{-\sysi}$, respectively. Analogously, Proposition~\ref{prop:TDI} manifests that the nonlinear system phase provides a valuable dynamic multiplier $\boldsymbol{\Psi}$ having nice physical interpretations. With the operator $\sysm={\cos\alpha \sysi - \sin\alpha \sysh}$ given in (\ref{eq:auxiliary_M}), the characterization of semi-sectorial systems in (\ref{eq:TDI}) is equivalent to that via the IQC defined by the multiplier $\boldsymbol{\Psi}=\tbt{0}{\boldsymbol{M}}{\boldsymbol{M}^*}{0}$. Therefore, the proof of the nonlinear small phase theorem can also be established by the IQC stability theorem \cite{Khong:21, Rantzer:97}.

Notice that the phase information of a system $\sysp$ is extracted and stored by the operator $\boldsymbol{M}$ given in (\ref{eq:auxiliary_M}).  This $\sysm$ can be represented in a more comprehensible manner in the frequency domain, namely, $M(j\omega)=\exp\rbkt{j\alpha\text{sgn}(\omega)}$. This special $\boldsymbol{M}$ itself deserves further discussion.
The study of this $\boldsymbol{M}$ as a spatial-domain filter in the image processing and optics disciplines dates back to the 1990s \cite{Lohmann:96}. If we regard  $\alpha$ here as a phasic parameter, the operator $\boldsymbol{M}$, called a \textit{fractional Hilbert transform filter} \cite{Lohmann:96,Lohmann:97, Davis:98,Venkitaraman:14}, is utilized to improve the performance for edge enhancement in image processing in lieu of the standard Hilbert transform filter $\sysh$. Specifically, the spatial-domain filter $\boldsymbol{M}$ with a designable $\alpha$ has the capability to decide where and to what degree edges of an input image are enhanced. In our case, by contrast, $\boldsymbol{M}$ presented in the phase supply rate (\ref{eq:supply}) is a time-domain filter. In addition, $\alpha$ is known as the intrinsic phasic quantity of $\sysp$. We believe that there exists a close connection between these two cases, and existing applications of the fractional Hilbert transform filter will facilitate our understanding of the phase supply rate.
\subsection{A Physical Interpretation of the Phase}
There is a nice physical interpretation of the nonlinear system phase in terms of the concept of energy. In Proposition~\ref{prop:TDI}, rewrite (\ref{eq:TDI}) in the following form: $\cos\alpha \ininf{u}{y }\geq \sin\alpha \ininf{
u}{\boldsymbol{H}y}$ for all $u \in \ltp$.
This inequality indicates an energy pair of a system $\sysp$, namely, the \textit{real} or \textit{active energy} $\ininf{u}{y}$ and the \textit{imaginary} or \textit{reactive energy} $\ininf{u}{\sysh y}$. As a quantity in nonlinear systems, the phase takes the role of balancing the real energy and reactive energy. To account for this, by setting $\beta\coloneqq \pi/2-\alpha \in \interval{-\pi/2}{3\pi/2}$ in Proposition~\ref{prop:TDI}, we have that, $\underline{\phi}(\sysp)\geq \beta-\pi$ and $\overline{\phi}(\sysp)\leq \beta$ if and only if,  $\sin\beta\ininf{u}{y}\geq \cos\beta \ininf{u}{\sysh y}$ for all $u \in \ltp$.
It follows that the phase infimum and phase supremum are connected with the ratio of the reactive energy to the real energy:
\begin{align*}
  \tan\overline{\phi}(\sysp) &= \sup_{u\in \ltp, \ininf{u}{y}\neq 0} \frac{\ininf{u}{\sysh y}}{\ininf{u}{y}}, \\
  \tan\underline{\phi}(\sysp) &= \inf_{u\in \ltp, \ininf{u}{y}\neq 0} \frac{\ininf{u}{\sysh y}}{\ininf{u}{y}}.
\end{align*}
Likewise, the phase supply rate (\ref{eq:supply}) has the physical meaning that this rate involves the instantaneous real power $u(t)^T y(t)$ and reactive power $u(t)^T (\boldsymbol{H}y)(t)$.

Recall that the passivity condition is concerned with the sign of energy $\ininf{u}{y}$, or, now more precisely, only the sign of the real energy. The phase quantity further reflects the potential influence of the reactive energy. Therefore, we believe that a better way to make full use of both the real and reactive energy information is to introduce the notion of phase in order to reduce conservatism in system analysis.

The concept of reactive or imaginary energy and power is not new; it exists and has different influences in various research fields. In circuit theory, the power triangle \cite[Section~11.6]{Alexander:12}, which shows the relationship between the real, reactive and complex power, is well known. In power system analysis, a similar idea of using the Hilbert transform to express reactive power is adopted in \cite{Chen:21, Nowomiejski:81, Cui:10}. In quantum mechanics, physical significance of the imaginary and complex energy is discussed in \cite[Section~134]{Landau:81}.

\subsection{The Multiplier Theorem}
We attempt to make a comparison between the famous multiplier theorem \cite{Zames:68}, \cite[Section~6.9]{Desoer:75} and the nonlinear small phase theorem. Generally speaking, the former is a common qualitative extension of the passivity theorem, while the latter is a quantitative one.

\begin{figure}[htb]
\centering
\setlength{\unitlength}{0.8mm}
\begin{picture}(70,25)
\thicklines
\put(0,20){\vector(1,0){8}} \put(10,20){\circle{4}}
\put(12,20){\vector(1,0){8}}
\put(20,15){\framebox(10,10){$\boldsymbol{P}$}}
\put(30,20){\vector(1,0){5}} \put(35,15){\framebox(10,10){$\boldsymbol{M}$}}
\put(45,20){\line(1,0){10}} \put(55,20){\vector(0,-1){13}}
{\thinlines
\multiput(18,13)(0,2){7}{\line(0,1){1}}
\multiput(18,13)(2,0){15}{\line(1,0){1}}
\multiput(47,13)(0,2){7}{\line(0,1){1}}
\multiput(18,27)(2,0){15}{\line(1,0){1}}
\multiput(18,-2)(0,2){7}{\line(0,1){1}}
\multiput(47,-2)(0,2){7}{\line(0,1){1}}
\multiput(18,-2)(2,0){15}{\line(1,0){1}}
\multiput(18,11)(2,0){15}{\line(1,0){1}}}
\put(35,5){\vector(-1,0){5}}
\put(55,5){\circle{4}}
\put(35,0){\framebox(10,10){$\boldsymbol{M}^{-1}$}}
\put(53,5){\vector(-1,0){8}}
 \put(20,0){\framebox(10,10){$\boldsymbol{C}$}}
\put(20,5){\line(-1,0){10}} \put(10,5){\vector(0,1){13}}
\put(5,10){\makebox(5,5){$y_2$}} \put(59,10){\makebox(5,5){$\boldsymbol{M}y_1$}}
\put(0,20){\makebox(5,5){$e_1$}} \put(60,-1){\makebox(5,5){$\boldsymbol{M}e_2$}}
\put(13,20){\makebox(5,5){$u_1$}} \put(30,0){\makebox(5,5){$u_2$}}
\put(10,10){\makebox(6,10){$-$}}
\put(65,5){\vector(-1,0){8}}
\end{picture}\caption{The multipliers $\boldsymbol{M}$ and $\boldsymbol{M}^{-1}$ are inserted into a feedback system.} \label{fig:multiplier1}
\end{figure}
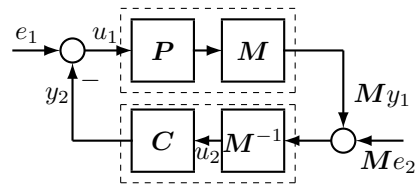

Let us revisit the stability analysis of $\gof$ in Fig.~\ref{fig:feedback}. The passivity theorem fails if $\sysp$ or $\sysc$ is not passive. Introducing multipliers can reduce this conservatism to a large extent, which is similar to the $\mulpi$-phase content in Section~\ref{sec:03}. Let a multiplier $\boldsymbol{M}\colon \lt \rightarrow \lt$ be a bounded linear operator with bounded inverse $\boldsymbol{M}^{-1}$. They are both inserted into the feedback loop, as is shown in Fig.~\ref{fig:multiplier1}. The fundamental idea is that, by doing this, the products $\boldsymbol{M}\sysp$ and $\sysc\boldsymbol{M}^{-1}$ can therefore satisfy the conditions of the passivity theorem. In addition, the stability of $\rbkt{\sysm\boldsymbol{P}}\,\#\,\rbkt{\boldsymbol{C}\sysm^{-1}}$ implies that of $\gof$, provided that $\boldsymbol{M}$ and $\boldsymbol{M}^{-1}$ are causal. Nevertheless, very often a noncausal $\boldsymbol{M}$ or $\boldsymbol{M}^{-1}$ is practically needed. The passivity theorem is still not applicable to $\rbkt{\sysm\boldsymbol{P}}\,\#\,\rbkt{\boldsymbol{C}\sysm^{-1}}$, since it requires causal components in the feedback loop.

\begin{figure}[htb]
\centering
\setlength{\unitlength}{0.8mm}
\begin{picture}(83,25)
\thicklines
\put(-3,20){\vector(1,0){8}} \put(7,20){\circle{4}}
\put(9,20){\vector(1,0){8}}
\put(17,15){\framebox(16,10){$(\boldsymbol{M}_-^{*})^{-1}$}}
\put(33,20){\vector(1,0){5}}
\put(38,15){\framebox(10,10){$\boldsymbol{P}$}}
\put(48,20){\vector(1,0){5}} \put(53,15){\framebox(10,10){$\boldsymbol{M}_+$}}
\put(63,20){\line(1,0){10}} \put(73,20){\vector(0,-1){13}}
{\thinlines
\multiput(15,13)(0,2){7}{\line(0,1){1}}
\multiput(15,13)(2,0){25}{\line(1,0){1}}
\multiput(65,13)(0,2){7}{\line(0,1){1}}
\multiput(15,27)(2,0){25}{\line(1,0){1}}
\multiput(15,-2)(0,2){7}{\line(0,1){1}}
\multiput(65,-2)(0,2){7}{\line(0,1){1}}
\multiput(15,-2)(2,0){25}{\line(1,0){1}}
\multiput(15,11)(2,0){25}{\line(1,0){1}}}
\put(38,5){\vector(-1,0){5}}
\put(73,5){\circle{4}}
\put(53,0){\framebox(10,10){$\boldsymbol{M}^{-1}_{+}$}}
\put(17,0){\framebox(16,10){$\boldsymbol{M}_-^{*}$}}
\put(71,5){\vector(-1,0){8}}
 \put(38,0){\framebox(10,10){$\boldsymbol{C}$}}
\put(17,5){\line(-1,0){10}}
\put(53,5){\vector(-1,0){5}}
\put(7,5){\vector(0,1){13}}
\put(-3,10){\makebox(5,5){$\boldsymbol{M}_-^{*}y_2$}} \put(78,10){\makebox(5,5){$\boldsymbol{M}_{+}y_1$}}
\put(-3,21){\makebox(5,5){$\boldsymbol{M}_-^* e_1$}} \put(78,-1){\makebox(5,5){$\boldsymbol{M}_+ e_2$}}
\put(33,20){\makebox(5,5){$u_1$}} \put(48,0){\makebox(5,5){$u_2$}}
\put(7,10){\makebox(6,10){$-$}}
\put(83,5){\vector(-1,0){8}}
\end{picture}\caption{A transformed system based on the factorization of the multipliers.} \label{fig:multiplier}
\end{figure}
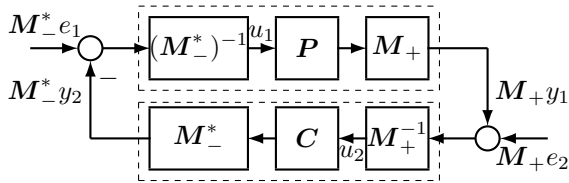

The multiplier theorem is developed under such circumstances, which indicates that $\boldsymbol{M}$ is required to meet a canonical factorization condition, i.e.,
\be \label{eq:factor}
\boldsymbol{M}=\boldsymbol{M}_- \boldsymbol{M}_+,
\ee
where $\boldsymbol{M}_-$ and $\boldsymbol{M}_+$ are invertible and $\boldsymbol{M}_+$, $\boldsymbol{M}_+^{-1}$, $\boldsymbol{M}_-^*$ and $(\boldsymbol{M}_-^{*})^{-1}$ are all causal and bounded. Owing to the artful factorization, all the components in the transformed system are causal, as is illustrated by Fig.~\ref{fig:multiplier}. In the sequel, the passivity theorem is applicable to the transformed system in Fig.~\ref{fig:multiplier}, and the stability of the transformed system implies that of $\gof$. Therefore, the key turns into seeking a suitable multiplier $\sysm$ which can be factorized in the form in (\ref{eq:factor}). In general, the multiplier theorem is often qualitative.

Proposition~\ref{prop:TDI} implies that the nonlinear small phase theorem involves an implicit bounded LTI multiplier $\boldsymbol{M}={\cos\alpha \sysi - \sin\alpha \sysh}$ in (\ref{eq:auxiliary_M}). Additionally, in Section~\ref{sec:dissipativity}, we know that $M(j\omega)=\exp\rbkt{j\alpha\text{sgn}(\omega)}$ and $M(s)\in\linf$. Indeed, this $\boldsymbol{M}$ is noncausal. As a consequence, if one wishes to re-establish the stability result in Theorem~\ref{Thm:nsp} using the multiplier theorem, a necessary step is to show that this $\boldsymbol{M}$ can be factorized in the form in (\ref{eq:factor}). Then, an open question arises: Does such a factorization exist for this $\sysm$?  In any case, unlike the multiplier theorem, we have no need of such a factorization in the nonlinear small phase theorem.

\section{Simulation Results}\label{sec:simu}
In this section, the nonlinear small phase theorem is demonstrated to be effective via a simulation study. Let $\sysp$ be a $2\times 2$ LTI system, with $P(s)$ given by
\bex P(s)=\tbt{\frac{s+6}{s^2+0.1s+1}}{\frac{0.2}{s^2+s+0.1}}{\frac{s+3}{s^2+4s+1}}{\frac{s+4}{s^2+s+1}}.
\eex
One can check that $P(s)$ is semi-sectorial, and compute that, $
\Phi(\sysp)= \interval{-{159.925}\pi/{180}}{{19.1142}\pi/{180}}$
and $\norm{\sysp}=60.8331$. Apparently, $\sysp$ is not passive and we calculate its negative passivity indices \cite{Vidyasagar:77} $\delta_1=\epsilon_1=-0.4526$.

Let $\sysc$ be a $2\times 2$ nonlinear system:
\begin{align*}
\tbo{\dot{x}_1(t)}{\dot{x}_2(t)}&=\tbo{-x_1(t)-x_2(t)-\rbkt{x_1(t)}^3}{-x_2(t)+x_1(t)-\rbkt{x_2(t)}^3}+\tbo{u_{2}^{(1)}(t)}{u_{2}^{(2)}(t)}\\
\tbo{y_{2}^{(1)}(t)}{y_{2}^{(2)}(t)}&=\tbo{x_1(t)}{x_2(t)}+\tbo{u_{2}^{(1)}(t)}{u_{2}^{(2)}(t)},
\end{align*}
with a zero initial condition $\obt{x_1(0)}{x_2(0)}^T=\obt{0}{0}^T$.
One can verify that $\sysc$ is stable with $\norm{\sysc}\in \interval{1.1}{2}$, which implies that the small gain theorem is inapplicable to the system $\gof$. More importantly, $\sysc$ is a very strictly passive system with passivity indices $\delta_2=2/3$ and $\epsilon_2=1/3$. Therefore, the index version of passivity theorem \cite{Vidyasagar:77} is also inapplicable since $\delta_1+\epsilon_2<0$. By Proposition~{\ref{lem:VSP}}, we know that the phase of $\sysc$ satisfies $\Phi(\sysc)\subseteq \interval{-19.4712\pi/180}{19.4712\pi/180}$.

\begin{figure}[htb]
  \centering
\includegraphics[width=3in]{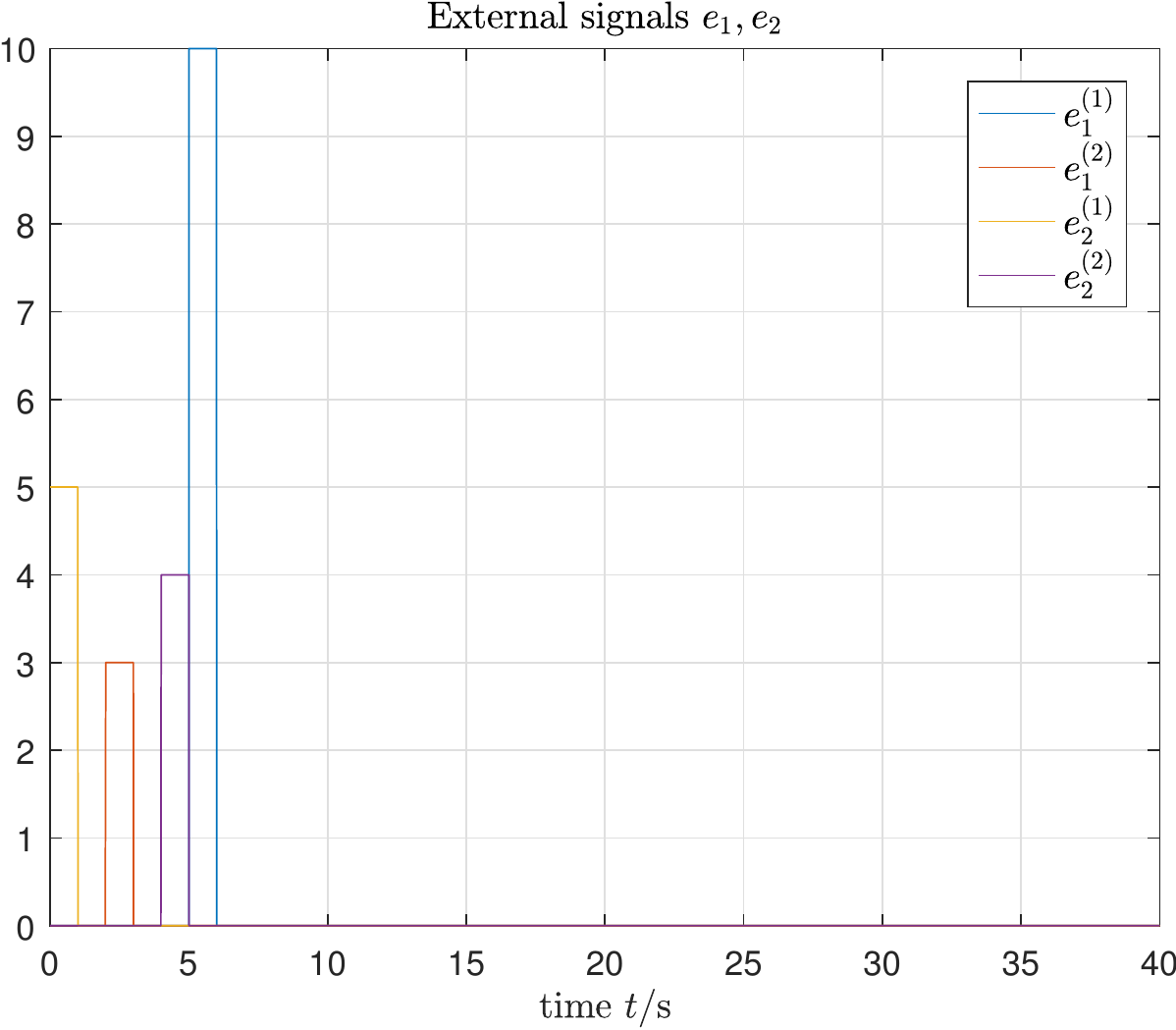}
\includegraphics[width=3in]{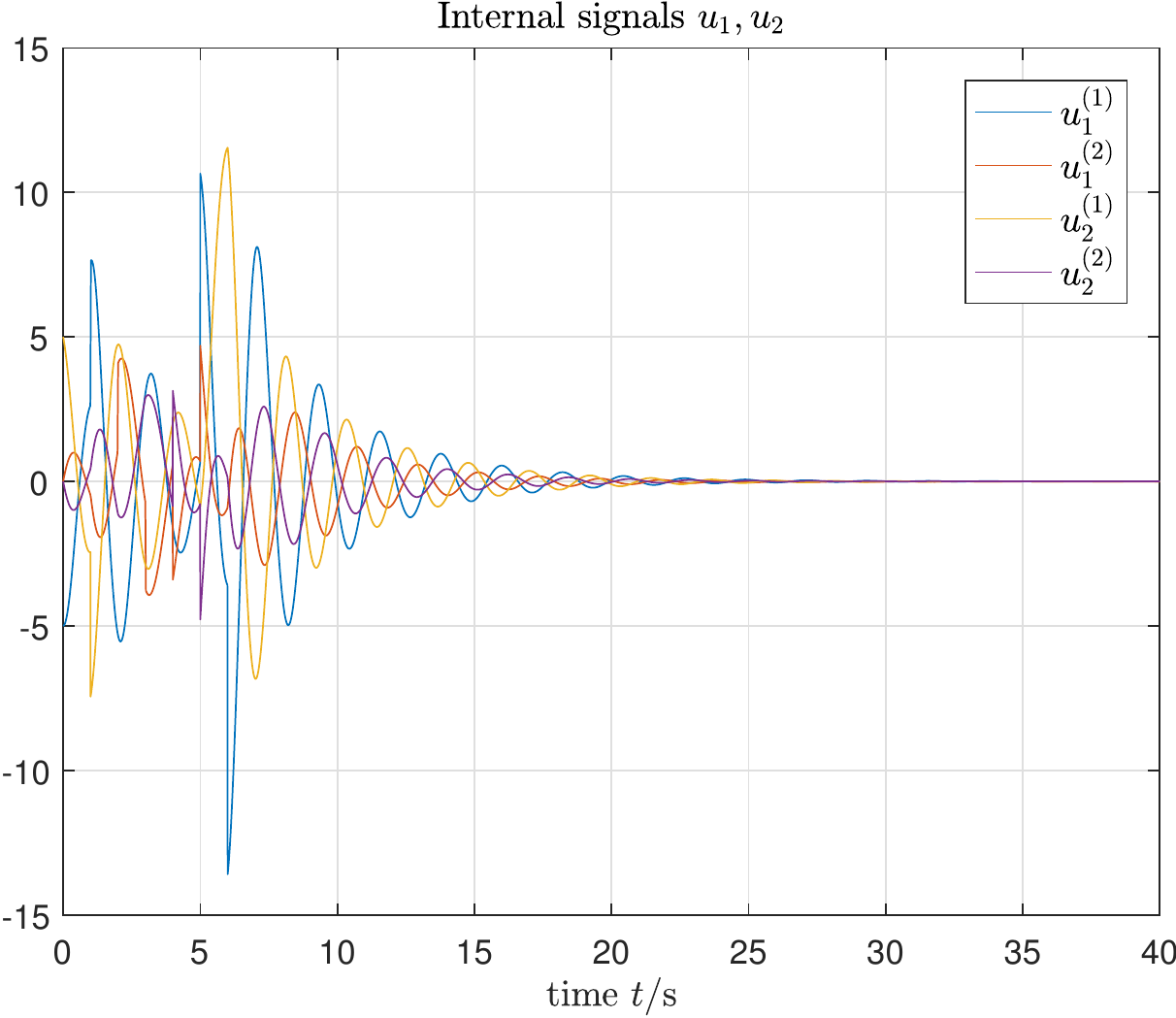}
\caption{The external signals $e_1, e_2$ (top) and internal signals $u_1$, $u_2$ (bottom) of the system $\gof$ in Fig.~\ref{fig:feedback}. Each external signal has two channels, and each channel consists of one rectangular pulse. All channels of the internal signals $u_1$, $u_2$ converge to zero in about thirty seconds.}
\label{fig:signals}
\end{figure}

In light of Theorem~\ref{Thm:nsp}, this $\gof$ is stable since the small phase condition (\ref{eq:small-phase}) is met.  As depicted in Fig.~\ref{fig:signals}, to test the stability, we adopt some rectangular pulses to generate the external signals $e_1$ and $e_2$ to stimulate the system $\gof$. The corresponding responses of the internal signals $u_1$ and $u_2$ converge to zero within thirty seconds. This reflects that the system $\gof$ is indeed stable.

\section{Conclusion}\label{sec:07}
In this paper, we take initiative to investigate the notion of nonlinear system phase. We define the phase for semi-sectorial systems from an input-output viewpoint and establish a corresponding nonlinear small phase theorem for closed-loop stability analysis. The proposed phase complements the $\lt$-gain and quantifies the passivity. In addition, the proposed theorem generalizes the MIMO LTI small phase theorem and passivity theorem in the literature. Subsequently, the proposed phase and theorem are further extended via the use of multipliers. Then, we connect the proposed theorem with the celebrated circle criterion for the Lur'e systems. We further reveal the relationship between the phase and existing notions. It shows that the phase can be understood through dynamic supply rates when interpreted via the dissipativity language, and through noncausal multipliers when embedded in the multiplier-based approach and IQC theory.

Three extensions beyond this work are under investigation. First, we are currently building a state-space theory for the nonlinear system phase. Second, we are studying a large-scale nonlinear network using the phase theory. Third, we are devoted to a mixed gain/phase theory for nonlinear systems. We look forward to bringing the phase onto the equal footing as the $\lt$-gain in nonlinear systems.

\appendix
In this appendix, we present all the proofs of theorems, propositions and corollaries in this paper, which are listed in order of appearance.

\begin{IEEEproof}[Proof of Proposition~\ref{lem:sector}]
  According to (\ref{eq:sector}), for all $0\neq u\in \ltp$ and for all $t\geq 0$, we have
  \bex
  (a+b)u(t)(\boldsymbol{N}u)(t) \geq ab\abs{u(t)}^2 + \abs{(\boldsymbol{N}u)(t)}^2.
  \eex
  Integrating both sides of the above inequality gives
  \be\label{eq:Lem1}
    (a+b) \ininf{u}{\boldsymbol{N}u}\geq ab\norm{u}^2_2 + \norm{\boldsymbol{N}u}^2_2.
  \ee
Recalling the properties of analytic signals (\ref{eq:analyticalproperty}), we have the following two identities: $2\norm{u_a}^2_2=\norm{u}^2_2$ and
$
\ininf{u}{\boldsymbol{N}u}= \rep\ininf{u+j\boldsymbol{H}u}{\boldsymbol{N}u}=2\rep{\ininf{u_a}{(\boldsymbol{N}u)_a}}$.
Substituting these two identities into (\ref{eq:Lem1}) and then dividing both sides of (\ref{eq:Lem1}) by $\norm{u_a}^2_2$ yields
\be\label{eq:Lem2}
(a+b) \rep{\frac{\ininf{u_a}{(\boldsymbol{N}u)_a}}{\norm{u_a}^2_2}} \geq ab +\frac{\norm{(\boldsymbol{N}u)_a}^2_2}{\norm{u_a}^2_2}.
\ee
Let $z\in \mathbb{C}$ be
$ z=z_R+jz_I\coloneqq{\ininf{u_a}{(\boldsymbol{N}u)_a}}/{\norm{u_a}^2_2}$.
Applying the Cauchy-Schwarz inequality to ${z}$ gives
\be\label{eq:Lem3}
\abs{z}=\frac{\abs{\ininf{u_a}{(\boldsymbol{N}u)_a}}}{\norm{u_a}^2_2}
\leq \frac{\norm{(\boldsymbol{N}u)_a}_2}{\norm{u_a}_2}.
\ee
Combining (\ref{eq:Lem2}) and (\ref{eq:Lem3}) shows that
$z$ satisfies
\bex
  \rbkt{a+b}z_R \geq ab + \rbkt{\frac{\norm{(\boldsymbol{N}u)_a}_2}{\norm{u_a}_2}}^2 \geq ab+\abs{z}^2=ab+z_R^2+z_I^2.
\eex
This gives
\bex
z_R^2-(a+b)z_R + \rbkt{\frac{a+b}{2}}^2 +z_I^2\leq \rbkt{\frac{b-a}{2}}^2.
\eex
Therefore, the set of $z$ satisfies $\abs{z-\rbkt{b+a}/{2}}\leq\rbkt{b-a}/{2}$,
which is exactly the disk $D(a, b)$ with the center $(b+a)/{2}$ and radius $(b-a)/2$. Noting (\ref{eq:analyticalproperty}), we have $\angle \ininf{u_a}{\boldsymbol{N}u}=\angle z$. It follows that $\angle \ininf{u_a}{\boldsymbol{N}u}\in \interval{\min \angle D(a, b)}{\max \angle D(a, b)}$. By the definition of phase, we conclude that \bex{\Phi}(\boldsymbol{N})\subseteq\interval{-{\arcsin\frac{b-a}{b+a}}}{\arcsin\frac{b-a}{b+a}} \subset \interval{-{\pi}/{2}}{{\pi}/{2}}.   \IEEEQEDhereeqn\eex
\end{IEEEproof}

\begin{IEEEproof}[Proof of Proposition~\ref{lem:VSP}]
 We use the same trick as in the proof of Proposition~\ref{lem:sector}. By utilizing the properties of analytic signals, we replace $\ininf{u}{\boldsymbol{P}u}$, $\norm{u}^2_2$ and  $\norm{\sysp u}^2_2$ in (\ref{eq:VSP_copy}) by $2\rep{\ininf{u_a}{(\boldsymbol{P}u)_a}}$, $2\norm{u_a}^2_2$ and $2\norm{(\boldsymbol{P}u)_a}^2_2$, respectively.
Then, dividing both sides of (\ref{eq:VSP_copy}) by $\norm{u_a}^2_2$ yields
\bex
\rep{\frac{\ininf{u_a}{(\boldsymbol{P}u)_a}}{\norm{u_a}^2_2}} \geq \delta  +\epsilon \frac{\norm{(\boldsymbol{P}u)_a}^2_2}{\norm{u_a}^2_2}.
\eex
We then follow the line of reasoning in the proof of Proposition~\ref{lem:sector}, and finally this will give
$\abs{z-{1}/\rbkt{2\epsilon}}^2\leq\rbkt{1-4\delta\epsilon}/\rbkt{4\epsilon^2}$,
where
$ z\coloneqq{\ininf{u_a}{(\boldsymbol{P}u)_a}}/
{\norm{u_a}^2_2}\in \mathbb{C}$.
Note that the constraint $\delta\epsilon\leq {1}/{4}$ always holds for a very strictly passive system \cite[Lemma 2.6]{Yu:13}. Therefore, we have
$
\abs{z-{1}/{(2\epsilon)}}\leq {\sqrt{1-4\delta\epsilon}}/{(2\epsilon)}$.
 Since $\angle \ininf{u_a}{\boldsymbol{P}u}=\angle z$, it follows that ${\Phi}(\boldsymbol{P})\subseteq\interval{-{\arcsin\sqrt{1-4\delta\epsilon}}}{\arcsin\sqrt{1-4\delta\epsilon}}.$
\end{IEEEproof}

\begin{IEEEproof}[Proof of Proposition~\ref{prop:LTI}]
First we show the subset relation $\subset$ of (\ref{eq:Prop_pf3}). For any $u \in \ltp$, using the Plancherel's theorem and (\ref{eq:freqH}), we have
\begin{align}\label{eq:Prop_pf1}
\ininf{u_a}{\sysp u}&=\ininf{\widehat{u_a}}{\widehat{\sysp u}}\notag\\
&=\frac{1}{2\pi}\int_{-\infty}^{\infty}\frac{1+\text{sgn}(\omega)}{2}\hat{u}(j\omega)^* {P}(j\omega)\hat{u}(\jw)d\omega. \notag\\
&=\frac{1}{2\pi}\int_{0}^{\infty}\hat{u}(j\omega)^* {P}(j\omega)\hat{u}(\jw)d\omega.
\end{align}
Following the definition of the Lebesgue integration, let $E_i$ be disjoint members of the $\sigma$-algebra of subsets of ${\mathbb{R}}_{+}$, and denote the simple function by $s_n(\jw)={1}/\rbkt{2\pi}\sum_{i=1}^{n} 1_{E_i} \hat{u}(j\omega_i)^* {P}(j\omega_i)\hat{u}(j\omega_i)$,
where $1_{E_i}$ denotes the indicator function and $\omega_i \in \interval{0}{\infty}$. Denote $\mu(E_i)$ as the Lebesgue measure on $E_i$. Integrating $s_n(\jw)$ from $0$ to $\infty$ yields
\begin{multline*}
   \int_{0}^{\infty}s_n(\jw)d\omega =\frac{1}{2\pi}\sum_{i=1}^{n}\mu(E_i)\hat{u}(j\omega_i)^* {P}(j\omega_i)\hat{u}(j\omega_i)\\
    \subset \text{conv}\bbkt{x^{*}{P}(\jw)x \in \mathbb{C} \mid \omega \in \interval{0}{\infty},  x \in \cn}.
\end{multline*}
It follows that
\begin{multline*}
  \frac{1}{2\pi}\int_{0}^{\infty}\hat{u}(j\omega)^* {P}(j\omega)\hat{u}(\jw)d\omega =\lim_{n\rightarrow\infty}\int_{0}^{\infty}s_n(\jw)d\omega\\
  \subset  \text{cl}~\text{conv}\bbkt{x^{*}{P}(\jw)x \in \mathbb{C} \mid \omega \in \interval{0}{\infty}, x \in \cn}.
\end{multline*}

Next we show the superset relation $\supset$ of (\ref{eq:Prop_pf3}). Given any $\omega_0\in\interval{0}{\infty}$ and $x\in\cn$, let $f\in \ltp$ with its $\hat{f}$ be chosen such that
\be\label{eq:Prop_pf2}
\abs{\hat{f}({\jw})}=\left\{
\begin{array}{l}
  c\quad \text{if}~\abs{\omega+\omega_0}<\epsilon_0~ \text{or}~\abs{\omega-\omega_0}<\epsilon_0 \\
  0\quad \text{otherwise},
\end{array}
\right.
\ee
where $\epsilon_0$ is a small positive number and $c$ is chosen so that $\hat{f}$ has a unit 2-norm, i.e., $c=\sqrt{{\pi}/({2\epsilon_0})}$. Let $u\coloneqq x f$. Then
\begin{align*}
 \ininf{u_a}{\syspu}&=\frac{1}{2\pi}\int_{0}^{\infty}\hat{u}(j\omega)^* {P}(j\omega)\hat{u}(\jw)d\omega \\
 &=\frac{1}{2\pi}\int_{0}^{\infty}\hat{f}(j\omega)^*x^*   {P}(j\omega)x\hat{f}(\jw)d\omega \\
 &\rightarrow x^{*}{P}(j\omega_0)x~\text{as}~\epsilon_0 \rightarrow 0.
\end{align*}
It follows that $x^{*}{P}(j\omega_0)x \in~\text{cl}\bbkt{W^{\prime}(\sysp)}$ for all $\omega_0 \in\interval{0}{\infty}$ and $x\in \cn$. Equivalently, for all $\omega \in\interval{0}{\infty}$, we have $\bbkt{x^{*}{P}(\jw)x \mid x \in \cn} \subset \text{cl}\bbkt{W^{\prime}(\sysp)}$. For a causal stable LTI system $\sysp$,  $\text{cl}\bbkt{W^{\prime}(\sysp)}$ is a closed convex cone by the Toeplitz-Hausdorff theorem in \cite[Chapter 17]{Halmos:74}. Hence we have proved the superset relation $\supset$ of (\ref{eq:Prop_pf3}).
\end{IEEEproof}

\begin{IEEEproof}[Proof of Proposition~\ref{prop:LTI_sectorial}]
First note that statement~(a) follows from statement~(c) by taking $\mulpi=\sysi$. Moreover, the proof of statement~(b) can be established via the same arguments as in  Proposition~\ref{prop:LTI}, and is omitted for simplicity. For statement~(c), we only show the sectorial system case, while the semi-sectorial system case can be proved by taking $\epsilon\rightarrow 0$ in the former. Sufficiency: By hypothesis, for all $u\in \ltp$ and all $\omega \in\interval{0}{\infty}$, we have
\bex
\hat{u}(\jw)^*\text{He}\rbkt{e^{j\alpha}\Pi(\jw)^*P(\jw)}\hat{u}(\jw)\geq2\epsilon \abs{P(\jw)\hat{u}(\jw)}^2.
\eex
Integrating both sides of the inequality from $0$ to $\infty$ gives
 \begin{multline*}
  \rep\sbkt{e^{j\alpha}\frac{1}{2\pi}\int_{0}^{\infty} \hat{u}(\jw)^*\Pi(\jw)^*P(\jw)\hat{u}(\jw)d\omega}\\
\geq  2\epsilon \frac{1}{2\pi}\int_{0}^{\infty} \abs{P(\jw)\hat{u}(\jw)}^2d\omega  = {\epsilon} \norm{P \hat{u}}_2^2.
\end{multline*}
According to the Plancherel's theorem and (\ref{eq:Prop_pf1}), we obtain
\begin{align*}
{\epsilon} \norm{\sysp u}_2^2 &\leq \rep\rbkt{e^{j\alpha}\ininf{u_a}{\mulpi^*\sysp u}}\\
&=  \cos\alpha\rep\ininf{\mulpi u_a}{\syspu} - \sin\alpha\imp\ininf{\mulpi u_a}{\syspu}.
\end{align*}
This implies that $\sysp$ is $\mulpi$-sectorial. Necessity: Suppose that $\sysp$ is $\mulpi$-sectorial. By (\ref{eq:SS}), there exist $\alpha \in \interval{-\pi}{\pi}$ and $\epsilon>0$ such that,
  $\rep\rbkt{e^{j\alpha} \ininf{\mulpi u_a}{\sysp u}} \geq \epsilon \norm{\sysp u}_2^2$ for all $u \in \ltp$.
Given any $\omega_0\in\interval{0}{\infty}$ and $x\in\cn$, we can adopt the same construction of $f\in \ltp$, parameterized by $\epsilon_0$, as in (\ref{eq:Prop_pf2}) in the proof of Proposition~\ref{prop:LTI}. Let $u=x f$, and then we have
  \begin{multline*}
     \rep\sbkt{e^{j\alpha}  \frac{1}{2\pi}\int_{0}^{\infty}\hat{f}(j\omega)^*x^* \Pi(\jw)^*  {P}(j\omega)x\hat{f}(\jw)d\omega}\\
       \geq  2\epsilon \frac{1}{2\pi}\int_{0}^{\infty}\abs{P(\jw)x\hat{f}(\jw)}^2 d\omega.
   \end{multline*}
  The above inequality approaches
   \begin{align*}
      \rep\rbkt{e^{j\alpha} x^{*}\Pi(\jw)^*{P}(j\omega_0)x}\geq 2\epsilon x^* P(\jw_0)^* P(\jw_0)x
  \end{align*}
  as $\epsilon_0 \rightarrow 0$. It follows that
  \bex
  x^* \sbkt{\text{He}\rbkt{e^{j\alpha}\Pi(\jw)^* P(\jw_0)}-2\epsilon P(\jw_0)^*P(\jw_0) } x \geq 0,
  \eex
  which gives that,  for all $\omega\in \interval{0}{\infty}$,
  \bex
  \text{He}\rbkt{e^{j\alpha} \Pi(\jw)^*P(\jw)}\geq 2\epsilon P(\jw)^*P(\jw).   \IEEEQEDhereeqn
  \eex
\end{IEEEproof}

\begin{IEEEproof}[Proof of Proposition~\ref{prop:freq_sectorial}]
By the conjugate symmetric and linear properties, $P(s)$ is frequency-wise sectorial if and only if, for each $\omega\in \interval{0}{\infty}$, the set $\bbkt{x^*P(\jw)x  \in \mathbb{C} \mid 0\neq x \in \cn}$ is contained in an open complex half-plane. This is also equivalent to the fact that, for all $\omega\in \interval{0}{\infty}$,  there exists a continuous conjugate symmetric function $\beta\colon \interval{-\infty}{\infty} \rightarrow \mathbb{C}$, i.e., $\beta(x)=\overline{\beta(-x)}$ for all $x\in \interval{-\infty}{\infty}$, with $\angle\beta\in \interval{-\pi}{\pi}$ such that $\bbkt{e^{j\angle \beta(\omega)}x^*P(\jw)x  \in \mathbb{C} \mid  0\neq x \in \cn}$ is contained in the open right half-plane. In other words, it holds that
\bex
\rep\rbkt{e^{j\angle\beta(\omega)}x^*P(\jw)x}>0,\quad \forall 0\neq x\in \mathbb{C}^n~\text{and}~\omega\in \interval{0}{\infty}.
\eex
Equivalently, there exists a sufficiently small $\delta>0$ such that $\text{He}(e^{j\angle\beta(\omega)}P(\jw))\geq \delta I$ for all $\omega\in \interval{0}{\infty}$, where $I$ is the identity matrix. We then define the multiplier $\Pi(\jw)=e^{-j\angle \beta(\omega)}\in \linf$. It follows that, for all $\omega\in \interval{0}{\infty}$, we have
\begin{align*}
&x^*\text{He}\rbkt{\Pi(\jw)^*P(\jw)} x=x^*\text{He}\rbkt{e^{j\angle\beta(\omega)}P(\jw)}x\\
\geq &  \delta x^*x \geq \frac{\delta}{\overline{\sigma}(P(\jw))^2} x^*P(\jw)^*P(\jw)x, \quad \forall 0\neq x\in \mathbb{C}^n,
\end{align*}
where $\overline{\sigma}(\cdot)$ denotes the largest singular value of a complex matrix. According to statement (c) of Proposition~\ref{prop:LTI_sectorial}, with $\alpha=0$, $P(s)$ is $\mulpi$-sectorial.
\end{IEEEproof}

\begin{IEEEproof}[Proof of Theorem~\ref{Thm:nsp}]
We use a homotopy method $\tau \in \interval{0}{1}$ and several steps to prove the result, which are inspired by \cite{Megretski:97}.
\begin{description}
  \item[{Step~1}:]\quad For all $u \in \ltp$ and $\tau \in \interval{0}{1}$, show that there exists $c_0>0$, independent of $\tau$, such that
      \bex
      \norm{u}_2\leq c_0  \norm{\tbt{\sysi}{\tau\sysc}{-\sysp}{\sysi} u}_2.\eex
\end{description}
Let $y_1=\sysp u_1$ and $y_2=\tau\sysc u_2$. The case $\tau=0$ is trivial since $\sysp$ is open-loop stable. Noting that $\Phi(\sysc)=\Phi(\tau\sysc)$ for all $\tau\in \interval[open left]{0}{1}$, let $\phi=\max\bbkt{\abs{{\overline{\phi}}(\sysp)}, \abs{{\overline{\phi}}(\tau\sysc)}, \abs{{\underline{\phi}}(\sysp)},\abs{{\underline{\phi}}(\tau\sysc)} }$.
Then define $\beta \coloneqq {{\phi}} - {\pi}/{2}$. By hypothesis, there exist a sufficiently small $\gamma > 0$ and $\alpha \in \bbkt{\beta + \gamma, -\beta - \gamma}$ such that $e^{j\alpha}W^{\prime}(\sysp) \subset \ccp$  and  $e^{-j\alpha}W^{\prime}(\tau\sysc) \subseteq \ccp$. Thus, by (\ref{eq:sectorial}) and (\ref{eq:SS}), there exists $\epsilon > 0$ such that, for all $u_1, u_2 \in \ltp$,
\begin{align}\label{eq:thm1_1}
\begin{aligned}
  e^{-j\alpha}\ininf{(u_2)_a}{y_2} + e^{j\alpha}\ininf{y_2}{(u_2)_a} &\geq 0,\\
 e^{j\alpha}\ininf{(u_1)_a}{y_1} + e^{-j\alpha}\ininf{y_1}{(u_1)_a} &\geq   \epsilon \norm{y_1}^2_2.
\end{aligned}
\end{align}
Summing both sides of the inequalities yields
\begin{align*}
 2\rep\bbkt{e^{j\alpha}\ininf{(u_1+y_2-y_2)_a}{y_1}+
 e^{-j\alpha}\ininf{(u_2-y_1+y_1)_a}{y_2}}\\
  \geq \epsilon \norm{y_1-u_2+u_2}^2_2.
\end{align*}
Rearranging the terms in the above inequality and discarding the positive term $\epsilon \norm{y_1-u_2}^2_2$ gives
\begin{align*}
 \epsilon\norm{u_2}^2_2 \leq & 2\rep\bbkt{e^{j\alpha}\ininf{(u_1+y_2)_a}{y_1}+e^{j\alpha}\ininf{y_2}{(u_2-y_1)_a}}\\
   & + 2\epsilon\ininf{u_2-y_1}{u_2}.
\end{align*}
Taking the absolute value on both sides of the above inequality and applying the triangle and Cauchy-Schwarz inequalities gives
\begin{align*}
\epsilon\norm{u_2}^2_2 \leq& 2\abs{e^{j\alpha}}\rbkt{\abs{\ininf{y_2}{(u_2-y_1)_a}}+\abs{\ininf{(u_1+y_2)_a}{y_1}}}\\
&+2\epsilon\abs{\ininf{u_2-y_1}{u_2}}\\
\leq&\sqrt{2}\rbkt{\norm{u_2-y_1}_2\norm{y_2}_2+\norm{u_1+y_2}_2\norm{y_1}_2}\\
&+2\epsilon\norm{u_2-y_1}_2\norm{u_2}_2.
\end{align*}
This implies
\begin{align}\label{eq:u2}
 &\norm{u_2}^2_2 \notag\\
 \leq&  {\frac{2\sqrt{2}}{\epsilon}}\norm{\tbt{\sysi}{\tau\sysc}{-\sysp}{\sysi}u}_2\norm{y}_2 +2\norm{\tbt{\sysi}{\tau\sysc}{-\sysp}{\sysi}u}_2\norm{u_2}_2\notag\\
\leq& \rbkt{{2+{\frac{2\sqrt{2}}{\epsilon}}
\norm{\tbt{\sysp}{0}{0}{\tau\sysc}}}}\norm{\tbt{\sysi}{\tau\sysc}{-\sysp}{\sysi}u}_2\norm{u}_2\notag\\
\leq& \rbkt{{2+{\frac{2\sqrt{2}}{\epsilon}}
\norm{\tbt{\sysp}{0}{0}{\sysc}}}}\norm{\tbt{\sysi}{\tau\sysc}{-\sysp}{\sysi}u}_2\norm{u}_2\notag\\
\eqqcolon& c_1\norm{\tbt{\sysi}{\tau\sysc}{-\sysp}{\sysi}u}_2\norm{u}_2,
\end{align}
where the inequalities are based on the following facts:
\begin{align*}
\norm{u_1}_2, \norm{u_2}_2&\leq \norm{u}_2\\
\norm{y_1}_2, \norm{y_2}_2&\leq \norm{y}_2\\
 \norm{y}_2\leq \norm{\tbt{\sysp}{0}{0}{\tau\sysc}}\norm{u}_2&\leq \norm{\tbt{\sysp}{0}{0}{\sysc}}\norm{u}_2\\
\norm{u_2-y_1}_2, \norm{u_1+y_2}_2&\leq\norm{\tbt{\sysi}{\tau\sysc}{-\sysp}{\sysi}u}_2.
\end{align*}
At the same time, using $y_2=\tau\boldsymbol{C}u_2$ gives
\bex\norm{y_2+u_1-u_1}^2_2 \leq \tau^2\norm{\boldsymbol{C}}^2\norm{u_2}^2_2.
\eex
After routine computations, we obtain
\begin{align}\label{eq:u1}
\norm{u_1}^2_2&\leq \tau^2\norm{\boldsymbol{C}}^2\norm{u_2}^2_2  +2\ininf{y_2+u_1}{u_1}\notag\\
&\leq \norm{\boldsymbol{C}}^2\norm{u_2}^2_2 +2\norm{y_2+u_1}_2\norm{u_1}_2\notag\\
&\leq 2\norm{\boldsymbol{C}}^2\norm{u_2}^2_2 +2 \norm{\tbt{\sysi}{\tau\sysc}{-\sysp}{\sysi}u}_2\norm{u}_2\notag\\
&\leq \sbkt{2+\norm{\boldsymbol{C}}^2c_1} \norm{\tbt{\boldsymbol{I}}{\tau\boldsymbol{C}}{-\boldsymbol{P}}{\boldsymbol{I}}u}_2\norm{u}_2.
\end{align}
Adding up (\ref{eq:u2}) and (\ref{eq:u1}) yields
\bex
\norm{u}_2\leq \sbkt{{2+\rbkt{1+\norm{\boldsymbol{C}}^2}c_1}}
\norm{\tbt{\boldsymbol{I}}{\tau\boldsymbol{C}}{-\boldsymbol{P}}{\boldsymbol{I}}u}_2.
\eex
Therefore, there exists a constant $c_0>0$ such that
\be\label{eq:thm1_bd}
\norm{u}_2 \leq c_0 \norm{\tbt{\boldsymbol{I}}{\tau\boldsymbol{C}}{-\boldsymbol{P}}{\boldsymbol{I}}u}_2
\ee
for all $u \in \ltp$ and $\tau \in \interval{0}{1}$.
\begin{description}
  \item[{Step~2}:]\quad Show that the stability of $\boldsymbol{P}\,\#\,\rbkt{\tau\boldsymbol{C}}$ implies the stability of $\boldsymbol{P}\,\#\,\sbkt{(\tau+\nu)\boldsymbol{C}}$ for all $\abs{\nu}< \mu= {1}/({c_0\norm{\boldsymbol{C}}})$ where $\mu$ is independent of $\tau$.
\end{description}
By the well-posedness assumption, the inverse  $\tbt{\boldsymbol{I}}{\tau\boldsymbol{C}}{-\boldsymbol{P}}{\boldsymbol{I}}^{-1}$ is well defined on $\ltep$. We assume that
 $\tbt{\boldsymbol{I}}{\tau\boldsymbol{C}}{-\boldsymbol{P}}{\boldsymbol{I}}^{-1}$ is bounded on $\ltp$. Given $u\in \ltep$, we define
\bex
u_T\coloneqq \tbt{\boldsymbol{I}}{\tau\boldsymbol{C}}{-\boldsymbol{P}}{\boldsymbol{I}}^{-1}\boldsymbol{\Gamma}_T\rbkt{\tbt{\boldsymbol{I}}{\tau\boldsymbol{C}}{-\boldsymbol{P}}{\boldsymbol{I}}u} \in \ltp.
\eex
Then we have
\begin{align*}
&  \norm{\boldsymbol{\Gamma}_T u}_2=\norm{\boldsymbol{\Gamma}_T u_T}_2\leq \norm{u_T}_2\\
   \leq & c_0\norm{\tbt{\boldsymbol{I}}{\tau\boldsymbol{C}}{-\boldsymbol{P}}{\boldsymbol{I}}u_T}_2=c_0\norm{\boldsymbol{\Gamma}_T\rbkt{\tbt{\boldsymbol{I}}{\tau\boldsymbol{C}}{-\boldsymbol{P}}{\boldsymbol{I}}u}}_2\\
  \leq & {c_0\norm{\boldsymbol{\Gamma}_T\rbkt{\tbt{\boldsymbol{I}}{(\tau+\nu)\boldsymbol{C}}{-\boldsymbol{P}}{\boldsymbol{I}}u}-\boldsymbol{\Gamma}_T\rbkt{\tbt{0}{\nu \boldsymbol{C}}{0}{0}u}}_2}\\
  \leq & c_0\norm{\boldsymbol{\Gamma}_T\rbkt{\tbt{\boldsymbol{I}}{(\tau+\nu)\boldsymbol{C}}{-\boldsymbol{P}}{\boldsymbol{I}}u}}_2+c_0\abs{\nu}\norm{\boldsymbol{C}}\norm{\boldsymbol{\Gamma}_T u}_2,
\end{align*}
where the last inequality uses the facts that $\sysc$ is causal and  $\norm{\boldsymbol{\Gamma}_T (\cdot)}_2$ is a nondecreasing function of $T$. This gives
\bex
 \norm{\boldsymbol{\Gamma}_T u}_2\leq
 \frac{c_0}{ \rbkt{1-c_0\abs{\nu}\norm{\boldsymbol{C}}}}\norm{\boldsymbol{\Gamma}_T\rbkt{\tbt{\boldsymbol{I}}{(\tau+\nu)\boldsymbol{C}}{-\boldsymbol{P}}{\boldsymbol{I}}u}}_2,
\eex
provided that  $\abs{\nu} < {1}/({c_0\norm{\boldsymbol{C}}})\eqqcolon\mu$.

\begin{description}
  \item[{Step~3}:]\quad Show that $\boldsymbol{P}\,\#\,\rbkt{\tau\boldsymbol{C}}$ is stable when $\tau=1$.
\end{description}
When $\tau=0$, $\tbt{\boldsymbol{I}}{\tau\boldsymbol{C}}{-\boldsymbol{P}}{\boldsymbol{I}}^{-1}$ is bounded since $\boldsymbol{P}$ is open-loop stable. It is shown in {Step~2} that  $\tbt{\boldsymbol{I}}{\tau\boldsymbol{C}}{-\boldsymbol{P}}{\boldsymbol{I}}^{-1}$ is bounded for $\tau < \mu$, then for $\tau < 2\mu$ using the iterative process, etc. By induction, $\tbt{\boldsymbol{I}}{\tau\boldsymbol{C}}{-\boldsymbol{P}}{\boldsymbol{I}}^{-1}$  is bounded for all $\tau \in \interval{0}{1}$. We conclude that $\gof$ is stable by setting $\tau=1$.
\end{IEEEproof}

\begin{IEEEproof}[Proof of Theorem~\ref{thm:gspt}]
  This theorem can be proven using the analogous arguments as in the proof of Theorem~\ref{Thm:nsp}, except for the following differences. First, as opposed to (\ref{eq:thm1_1}),
  there exists $\epsilon > 0$ such that, for all $u_1, u_2 \in \ltp$,
\begin{align*}
  e^{-j\alpha}\ininf{\mulpi^*(u_2)_a}{y_2} + e^{j\alpha}\ininf{y_2}{\mulpi^*(u_2)_a} &\geq 0,\\
 e^{j\alpha}\ininf{\mulpi(u_1)_a}{y_1} + e^{-j\alpha}\ininf{y_1}{\mulpi(u_1)_a} &\geq   \epsilon \norm{y_1}^2_2.
\end{align*}
Second, note that the following term is exactly cancelled:
 \bex
\ininf{-\mulpi(y_2)_a}{y_1}+\ininf{y_2}{\mulpi^*(y_1)_a}=0,
\eex
which can be easily shown in the frequency domain by the Plancherel's theorem and properties of the Hilbert transform. We then follow the line of reasoning in the proof of Theorem~\ref{Thm:nsp}, and this will give us the new constant
\bex
c_1=2+\frac{2\sqrt{2}\norm{\mulpi}}{\epsilon}\norm{\tbt{\sysp}{0}{0}{\sysc}}
\eex
in lieu of that in (\ref{eq:u2}). The constants $c_2$ in (\ref{eq:u1}) and $c_0$ in (\ref{eq:thm1_bd}) can be derived accordingly.
\end{IEEEproof}

\begin{IEEEproof}[Proof of Corollary~\ref{coro:LTI}]
The proof can be established by the combination of Proposition~\ref{prop:LTI}, statement (a) of  Proposition~\ref{prop:LTI_sectorial} and Theorem~\ref{Thm:nsp}.
\end{IEEEproof}

\begin{IEEEproof}[Proof of Corollary~\ref{coro:freq_LTI}]
We choose the multiplier $\Pi(\jw)=e^{-j\angle\beta(\omega)}\in \linf$ which has been constructed in the proof of Proposition~\ref{prop:freq_sectorial}.
 Note that, for a frequency-wise sectorial system $P(s)$, the angular numerical range of the matrices $P(\jw)$ at different frequencies $\omega$ may be contained in different half-planes. Then, using this $\Pi(\jw)$, we make the angular numerical ranges of the matrices $\Pi(\jw)^*P(\jw)$ at all frequencies $\omega$ contained in the same half-plane. In other words, we make $\Pi(s)^*P(s)$  sectorial and $P(s)$ $\mulpi$-sectorial, as elaborated in the proof of Proposition~\ref{prop:freq_sectorial}. The similar argument applies to a frequency-wise semi-sectorial system $C(s)$.  The proof thereupon is completed by invoking Theorem~\ref{thm:gspt}.
\end{IEEEproof}

\begin{IEEEproof}[Proof of Corollary~\ref{coro:lure}]
Notice that $\Phi(\tau P(s))=\Phi(P(s))$ for all $\tau>0$. Therefore, for the semi-sectorial $\tau P(s)$, by assumption (\ref{eq:coro2}), the two extreme cases of $\Phi(\tau P(s))$ will be
 \begin{align*}
  \Phi(\tau P(s))&\subseteq\interval[open right]{{-\arcsin\frac{b-a}{b+a}}}{\pi-\arcsin\frac{b-a}{b+a}},\\
  \Phi(\tau P(s))&\subseteq\interval[open left]{{\arcsin\frac{b-a}{b+a}-\pi}}{\arcsin\frac{b-a}{b+a}}.
 \end{align*}
Additionally, on the basis of Proposition~\ref{lem:sector}, ${\Phi}(\boldsymbol{N})$ satisfies (\ref{eq:phase_sector}). Thus, the small phase condition (\ref{eq:small-phase}) is satisfied in either case and the proof is completed by invoking Theorem~\ref{Thm:nsp}.
\end{IEEEproof}

\begin{IEEEproof}[Proof of Proposition~\ref{prop:parallel}]
It suffices to show that, for any $\sysp_1, \sysp_2 \in \mathcal{P}(\alpha, \beta)$ and $\theta_1, \theta_2>0$, we have $\sysp\coloneqq \theta_1\sysp_1+\theta_2\sysp_2\in \mathcal{P}(\alpha,\beta)$. We first consider a $\pi$-length interval $\interval{\beta-\pi}{\beta}$. Clearly, it holds that $\interval{\alpha}{\beta}\subseteq \interval{\beta-\pi}{\beta}$. Since $\sysp_1$ and $\sysp_2$ are sectorial, so are $\theta_1\sysp_1$ and $\theta_2\sysp_2$. Then, for all $u\in \ltp$, we have
\begin{align*}
  & \cos\rbkt{{\pi}/{2}-\beta}\rep\ininf{u_a}{\syspu} - \sin\rbkt{{\pi}/{2}-\beta}\imp\ininf{u_a}{\syspu}\\
= &\cos\rbkt{{\pi}/{2}-\beta}\rbkt{\rep\ininf{u_a}{\theta_1\sysp_1u}+\rep\ininf{u_a}{\theta_2\sysp_2u}}\\
 &- \sin\rbkt{{\pi}/{2}-\beta}\rbkt{\imp\ininf{u_a}{\theta_1\sysp_1u}+\imp\ininf{u_a}{\theta_2\sysp_2u}}\\
   \geq & \epsilon_1\norm{\theta_1\sysp_1 u}_2^2+\epsilon_2\norm{\theta_2\sysp_2 u}_2^2 \geq \epsilon\norm{\sysp u}_2^2,
\end{align*}
where $\epsilon=\min (\epsilon_1, \epsilon_2)/2 >0$ and the first inequality is due to definition~(\ref{eq:SS}) for sectorial systems, and the last inequality uses the fact that, for all $v, f\in\ltp$,  $\norm{v-f}^2_2+\norm{f}^2_2 \geq \norm{v}_2^2/2$. Then, according to~(\ref{eq:SS}), $\sysp$ is sectorial and $\Phi(\sysp)\subseteq \interval{\beta-\pi}{\beta}$. Moreover, note that $\interval{\alpha}{\beta}$ is contained in another $\pi$-length interval $\interval{\alpha}{\alpha+{\pi}}$. Following the same arguments
as above, we can derive $\Phi(\sysp)\subseteq \interval{\alpha}{\alpha+\pi}$. Intersecting the two $\pi$-length intervals yields $\Phi(\sysp)\subseteq \interval{\alpha}{\beta}$, which means $\sysp\in\mathcal{P}(\alpha,\beta)$.\hfill \IEEEQEDhere
\end{IEEEproof}

\begin{IEEEproof}[Proof of Proposition~\ref{prop:CL_phase}]
We only prove the inclusion relation of $\Phi(\boldsymbol{G}_1)$ since that of $\Phi(\boldsymbol{G}_2)$ can be shown in an analogous way.
The stability of $\gof$ implies that $\boldsymbol{G}_1$ is stable. Since $e_2=0$, according to the well-posedness $u_1=e_1-y_2$ and $u_2=y_1$, for all $e_1 \in \ltp$, we have
\begin{align*}
\ininf{e_{1a}}{y_1}&=\ininf{u_{1a}+y_{2a}}{y_1}=\ininf{u_{1a}}{y_1}+\ininf{y_{2a}}{u_2}\\
& = \ininf{u_{1a}}{y_1}+\overline{\ininf{u_2}{y_{2a}}}=\ininf{u_{1a}}{y_1}+\overline{\ininf{u_{2a}}{y_{2}}}.
\end{align*}
This gives  \bex\angle \ininf{e_{1a}}{y_1} = \angle\sbkt{ \ininf{u_{1a}}{y_1}+\overline{\ininf{u_{2a}}{y_{2}}}},\eex
with $\angle \ininf{u_{1a}}{y_1}\in \interval{\underline{\phi}(\sysp)}{\overline{\phi}(\sysp)}$ and $\angle \overline{\ininf{u_{2a}}{y_{2}}}\in \interval{-\overline{\phi}(\sysc)}{-\underline{\phi}(\sysc)}$.  Note that, for any $0\neq a,b\in \mathbb{C}$, we have
\be\label{eq:scalar_ineq}
\min \bbkt{\angle a, \angle b} \leq \angle (a+b) \leq \max \bbkt{\angle a, \angle b}
\ee
when $\abs{\angle a -\angle b}\leq \pi$, except for the case $a+b=0$ where $\angle (a+b)$ is undefined. By hypothesis, the following inequalities
\begin{align*}
  -\pi& \leq {\underline{\phi}}(\sysp) + {\underline{\phi}}(\sysc) \leq {\overline{\phi}}(\sysp) + {\underline{\phi}}(\sysc) \leq  {\overline{\phi}}(\sysp) +  {\overline{\phi}}(\sysc) \leq \pi, \\
    -\pi& \leq {\underline{\phi}}(\sysp) + {\underline{\phi}}(\sysc) \leq {\underline{\phi}}(\sysp) +  {\overline{\phi}}(\sysc) \leq  {\overline{\phi}}(\sysp) +  {\overline{\phi}}(\sysc) \leq\pi
\end{align*}
hold. This guarantees that
\begin{align*}
  \abs{{\overline{\phi}}(\sysp) - \rbkt{-{\underline{\phi}}(\sysc)}}&\leq \pi\quad\text{and}\quad   \abs{{\underline{\phi}}(\sysp) - \rbkt{-{\overline{\phi}}(\sysc)}}\leq \pi,\\
    \abs{{\overline{\phi}}(\sysp) - \rbkt{-{\overline{\phi}}(\sysc)}}&\leq \pi \quad\text{and}\quad   \abs{{\underline{\phi}}(\sysp) - \rbkt{-{\underline{\phi}}(\sysc)}}\leq \pi.
\end{align*}
By applying (\ref{eq:scalar_ineq}), we obtain
      \bex
  \min\bbkt{\underline{\phi}(\sysp),-\overline{\phi}(\sysc)} \leq   \angle \ininf{e_{1a}}{y_1}\leq \max\bbkt{\overline{\phi}(\sysp),-\underline{\phi}(\sysc)}
   \eex
for all $e_1\in \ltp$, except for the case that $\ininf{e_{1a}}{y_1}=0$ for some $e_1$ where $\angle \ininf{e_{1a}}{y_1}$ is undefined. By the definition of phase, we conclude that
   $
\Phi(\boldsymbol{G}_1) \subseteq \interval {\min\bbkt{\underline{\phi}(\sysp),-\overline{\phi}(\sysc)}}{\max\bbkt{\overline{\phi}(\sysp),-\underline{\phi}(\sysc)}}$.
\end{IEEEproof}

\begin{IEEEproof}[Proof of Proposition~\ref{prop:TDI}]
Using (\ref{eq:sectorial}) and the properties (\ref{eq:analyticalproperty}), the condition $\Phi(\boldsymbol{P}) \subseteq \interval{-{\pi}/{2}-\alpha}{{\pi}/{2}-\alpha}$ is equivalent to
  \begin{align*}
    &\cos\alpha\rep\ininf{u_a}{\sysp u} - \sin\alpha\imp\ininf{u_a}{\sysp u}\\
    =& \cos\alpha\rep\ininf{u}{(\sysp u)_a} - \sin\alpha\imp\ininf{u}{(\sysp u)_a}\\
    =&{1}/{2}\cos\alpha\ininf{u}{\sysp u}-1/2\sin\alpha\ininf{u}{\sysh \sysp u}\geq 0
  \end{align*}
for all $u\in \ltp$. This is equivalent to saying that $
\ininf{u}{\cos\alpha \sysp u - \sin\alpha \boldsymbol{H}\sysp u}\geq 0$ for all $u \in \ltp$.
\end{IEEEproof}
\begin{IEEEproof}[Proof of Proposition~\ref{co:2}]
The proof follows directly from Proposition~\ref{prop:TDI} and the definition of ultimate dissipativity with respect to the dynamic $\rbkt{\boldsymbol{Q}, \boldsymbol{S}, \boldsymbol{R}}$-supply rate.
\end{IEEEproof}

\section*{Acknowledgment}
The authors would like to thank Dan Wang, Axel Ringh and Xin Mao of The Hong Kong University
of Science and Technology for useful discussions.

\bibliographystyle{IEEEtran}
\bibliography{Phase_arXiv_V2}

\begin{thebibliography}{10}
\providecommand{\url}[1]{#1}
\csname url@samestyle\endcsname
\providecommand{\newblock}{\relax}
\providecommand{\bibinfo}[2]{#2}
\providecommand{\BIBentrySTDinterwordspacing}{\spaceskip=0pt\relax}
\providecommand{\BIBentryALTinterwordstretchfactor}{4}
\providecommand{\BIBentryALTinterwordspacing}{\spaceskip=\fontdimen2\font plus
\BIBentryALTinterwordstretchfactor\fontdimen3\font minus
  \fontdimen4\font\relax}
\providecommand{\BIBforeignlanguage}[2]{{%
\expandafter\ifx\csname l@#1\endcsname\relax
\typeout{** WARNING: IEEEtran.bst: No hyphenation pattern has been}%
\typeout{** loaded for the language `#1'. Using the pattern for}%
\typeout{** the default language instead.}%
\else
\language=\csname l@#1\endcsname
\fi
#2}}
\providecommand{\BIBdecl}{\relax}
\BIBdecl

\bibitem{Postlethwaite:81}
I.~Postlethwaite, J.~Edmunds, and A.~MacFarlane, ``Principal gains and
  principal phases in the analysis of linear multivariable feedback systems,''
  \emph{IEEE Trans. Autom. Control}, vol.~26, no.~1, pp. 32--46, 1981.

\bibitem{Owens:84}
D.~Owens, ``The numerical range: {A} tool for robust stability studies?''
  \emph{Syst. \& Control Lett.}, vol.~5, no.~3, pp. 153--158, 1984.

\bibitem{Tits:99}
A.~L. Tits, V.~Balakrishnan, and L.~Lee, ``Robustness under bounded uncertainty
  with phase information,'' \emph{IEEE Trans. Autom. Control}, vol.~44, no.~1,
  pp. 50--65, 1999.

\bibitem{Laib:17}
K.~{Laib}, A.~{Korniienko}, M.~{Dinh}, G.~{Scorletti}, and F.~{Morel},
  ``Hierarchical robust performance analysis of uncertain large scale
  systems,'' \emph{IEEE Trans. Autom. Control}, vol.~63, no.~7, pp. 2075--2090,
  2018.

\bibitem{Chen:19}
W.~Chen, D.~Wang, S.~Z. Khong, and L.~Qiu, ``Phase analysis of {MIMO} {LTI}
  systems,'' in \emph{Proc. 58th IEEE Conf. Decision and Control}, Nice,
  France, 2019, pp. 6062--6067.

\bibitem{Chen:21}
------, ``A phase theory of {MIMO} {LTI} systems,'' \emph{Manuscript Submitted
  for Publication}, 2021.

\bibitem{Zames:66}
G.~Zames, ``On the input-output stability of time-varying nonlinear feedback
  systems {P}art {\uppercase\expandafter{\romannumeral1}}: Conditions derived
  using concepts of loop gain, conicity, and positivity,'' \emph{IEEE Trans.
  Autom. Control}, vol.~11, no.~2, pp. 228--238, 1966.

\bibitem{Van:17}
A.~van~der Schaft, \emph{$L_2$-Gain and Passivity Techniques in Nonlinear
  Control}, 3rd~ed.\hskip 1em plus 0.5em minus 0.4em\relax Cham, Switzerland:
  Springer International Publishing AG, 2017.

\bibitem{Vidyasagar:93}
M.~Vidyasagar, \emph{Nonlinear Systems Analysis}, 2nd~ed.\hskip 1em plus 0.5em
  minus 0.4em\relax Englewood Cliffs, NJ: Prentice-Hall, 1993.

\bibitem{Anders:19}
A.~Rantzer, ``Lecture notes on {N}onlinear {C}ontrol and {S}ervo {S}ystems,''
  http://www.control.lth.se/education/engineering-program/frtn05-nonlinear-control-and-servo-systems,
  2019, accessed February 10, 2020.

\bibitem{Angeli:06}
D.~Angeli, ``Systems with counterclockwise input-output dynamics,'' \emph{IEEE
  Trans. Autom. Control}, vol.~51, no.~7, pp. 1130--1143, 2006.

\bibitem{Petersen:10}
I.~R. Petersen and A.~Lanzon, ``Feedback control of negative-imaginary
  systems,'' \emph{IEEE Control Systems Magazine}, vol.~30, no.~5, pp. 54--72,
  2010.

\bibitem{King:09}
F.~W. King, \emph{Hilbert Transforms}.\hskip 1em plus 0.5em minus 0.4em\relax
  New York, NY: Cambridge University Press, 2009.

\bibitem{Hahn:96}
S.~L. Hahn, \emph{Hilbert Transforms in Signal Processing}.\hskip 1em plus
  0.5em minus 0.4em\relax Norwood, MA: Artech House, 1996.

\bibitem{Mao:20}
X.~Mao, W.~Chen, and L.~Qiu, ``Phase analysis for discrete-time {LTI}
  multivariable systems,'' in \emph{Preprints of the 21st IFAC World Congress
  (Virtual)}, Berlin, Germany, 2020, pp. 4468--4473.

\bibitem{Wang:20}
D.~Wang, W.~Chen, S.~Z. Khong, and L.~Qiu, ``On the phases of a complex
  matrix,'' \emph{Linear Algebra Appl.}, vol. 593, pp. 152--179, 2020.

\bibitem{Lohmann:96}
A.~W. Lohmann, D.~Mendlovic, and Z.~Zalevsky, ``Fractional {H}ilbert
  transform,'' \emph{Opt. Lett.}, vol.~21, no.~4, pp. 281--283, 1996.

\bibitem{Lohmann:97}
A.~W. Lohmann, E.~Tepichin, and J.~Ramirez, ``Optical implementation of the
  fractional {H}ilbert transform for two-dimensional objects,'' \emph{Appl.
  Opt.}, vol.~36, no.~26, pp. 6620--6626, 1997.

\bibitem{Davis:98}
J.~A. Davis, D.~E. McNamara, and D.~M. Cottrell, ``Analysis of the fractional
  {H}ilbert transform,'' \emph{Appl. Opt.}, vol.~37, no.~29, pp. 6911--6913,
  1998.

\bibitem{Venkitaraman:14}
A.~Venkitaraman and C.~S. Seelamantula, ``Fractional {H}ilbert transform
  extensions and associated analytic signal construction,'' \emph{Signal
  Process.}, vol.~94, pp. 359--372, 2014.

\bibitem{Khalil:02}
H.~K. Khalil, \emph{Nonlinear Systems}, 3rd~ed.\hskip 1em plus 0.5em minus
  0.4em\relax Upper Saddle River, NJ: Prentice Hall, 2002.

\bibitem{Willems:72}
J.~C. Willems, ``Dissipative dynamical systems {P}art
  {\uppercase\expandafter{\romannumeral1}}: {G}eneral theory,'' \emph{Arch.
  Ration. Mech. Anal.}, vol.~45, no.~5, pp. 321--351, 1972.

\bibitem{Hill:80}
D.~J. Hill and P.~J. Moylan, ``Dissipative dynamical systems: {B}asic
  input-output and state properties,'' \emph{J. Franklin Inst.}, vol. 309,
  no.~5, pp. 327--357, 1980.

\bibitem{Megretski:97}
A.~Megretski and A.~Rantzer, ``System analysis via integral quadratic
  constraints,'' \emph{IEEE Trans. Autom. Control}, vol.~42, no.~6, pp.
  819--830, 1997.

\bibitem{Cho:68}
Y.-S. Cho and K.~S. Narendra, ``Stability of nonlinear time-varying feedback
  systems,'' \emph{Automatica}, vol.~4, no.~5, pp. 309--322, 1968.

\bibitem{Vidyasagar:77}
M.~Vidyasagar, ``${L}_2$-stability of interconnected systems using a
  reformulation of the passivity theorem,'' \emph{IEEE Trans. Circuits Syst.},
  vol.~24, no.~11, pp. 637--645, 1977.

\bibitem{Zames:68}
G.~Zames and P.~L. Falb, ``Stability conditions for systems with monotone and
  slope-restricted nonlinearities,'' \emph{SIAM J. Control}, vol.~6, no.~1, pp.
  89--108, 1968.

\bibitem{Gabor:46}
D.~Gabor, ``Theory of communication {P}art
  {\uppercase\expandafter{\romannumeral1}}: The analysis of information,''
  \emph{J. Inst. Electr. Eng. 3}, vol.~93, pp. 429--441, 1946.

\bibitem{Narendra:66}
K.~S. {Narendra} and P.~G. {Gallman}, ``An iterative method for the
  identification of nonlinear systems using a {H}ammerstein model,'' \emph{IEEE
  Trans. Autom. Control}, vol.~11, no.~3, pp. 546--550, 1966.

\bibitem{Desoer:75}
C.~A. Desoer and M.~Vidyasagar, \emph{Feedback Systems: {I}nput-Output
  Properties}.\hskip 1em plus 0.5em minus 0.4em\relax New York, NY: Academic
  Press, 1975.

\bibitem{Fu:05_2}
M.~Fu and L.~Xie, ``The sector bound approach to quantized feedback control,''
  \emph{IEEE Trans. Autom. Control}, vol.~50, no.~11, pp. 1698--1711, 2005.

\bibitem{Yu:13}
H.~Yu, F.~Zhu, M.~Xia, and P.~J. Antsaklis, ``Robust stabilizing output
  feedback nonlinear model predictive control by using passivity and
  dissipativity,'' in \emph{Proc. 2013 European Control Conf.}, 2013, pp.
  2050--2055.

\bibitem{Willems:71}
J.~C. Willems, \emph{The Analysis of Feedback Systems}.\hskip 1em plus 0.5em
  minus 0.4em\relax London, UK: The MIT Press, 1971.

\bibitem{Sandberg:64}
I.~W. Sandberg, ``A frequency-domain condition for the stability of feedback
  systems containing a single time-varying nonlinear element,'' \emph{Bell
  System Technical Journal}, vol.~43, no.~4, pp. 1601--1608, 1964.

\bibitem{Zames:66_2}
G.~Zames, ``On the input-output stability of time-varying nonlinear feedback
  systems {P}art {\uppercase\expandafter{\romannumeral2}}: Conditions involving
  circles in the frequency plane and sector nonlinearities,'' \emph{IEEE Trans.
  Autom. Control}, vol.~11, no.~3, pp. 465--476, 1966.

\bibitem{Willems:98}
J.~C. Willems and H.~L. Trentelman, ``On quadratic differential forms,''
  \emph{SIAM J. Control Optim.}, vol.~36, no.~5, pp. 1703--1749, 1998.

\bibitem{Willems:02}
------, ``Synthesis of dissipative systems using quadratic differential forms:
  Part~{\uppercase\expandafter{\romannumeral1}},'' \emph{IEEE Trans. Autom.
  Control}, vol.~47, no.~1, pp. 53--69, 2002.

\bibitem{Forni:13}
F.~Forni and R.~Sepulchre, ``On differentially dissipative dynamical systems,''
  \emph{IFAC Proceedings Volumes}, vol.~46, no.~23, pp. 15--20, 2013.

\bibitem{Van:13}
A.~J. van~der Schaft, ``On differential passivity,'' \emph{IFAC Proceedings
  Volumes}, vol.~46, no.~23, pp. 21--25, 2013.

\bibitem{Arcak:16}
M.~Arcak, C.~Meissen, and A.~Packard, \emph{Networks of Dissipative Systems:
  Compositional Certification of Stability, Performance, and Safety}.\hskip 1em
  plus 0.5em minus 0.4em\relax Cham, Switzerland: Springer International
  Publishing AG, 2016.

\bibitem{Seiler:15}
P.~Seiler, ``Stability analysis with dissipation inequalities and integral
  quadratic constraints,'' \emph{IEEE Trans. Autom. Control}, vol.~60, no.~6,
  pp. 1704--1709, 2015.

\bibitem{Scherer:18}
C.~W. Scherer and J.~Veenman, ``Stability analysis by dynamic dissipation
  inequalities: {O}n merging frequency-domain techniques with time-domain
  conditions,'' \emph{Syst. \& Control Lett.}, vol. 121, pp. 7--15, 2018.

\bibitem{Khong:21}
\BIBentryALTinterwordspacing
S.~Z. Khong, ``On integral quadratic constraints,'' \emph{IEEE Trans. Autom.
  Control (Early Access)}, 2021. [Online]. Available:
  \url{https://doi.org/10.1109/TAC.2021.3069665}
\BIBentrySTDinterwordspacing

\bibitem{Rantzer:97}
A.~Rantzer and A.~Megretski, \emph{System Analysis via {I}ntegral {Q}uadratic
  {C}onstraints {P}art~{\uppercase\expandafter{\romannumeral2}}: {A}bstract
  Theory}, ser. Technical Reports TFRT-7559.\hskip 1em plus 0.5em minus
  0.4em\relax Department of Automatic Control, Lund Institute of Technology,
  1997.

\bibitem{Alexander:12}
C.~K. Alexander and M.~N.~O. Sadiku, \emph{Fundamentals of Electric Circuits},
  5th~ed.\hskip 1em plus 0.5em minus 0.4em\relax New York, NY: McGraw-Hill
  Education, 2012.

\bibitem{Nowomiejski:81}
Z.~Nowomiejski, ``Generalized theory of electric power,'' \emph{Archiv f{\"u}r
  Elektrotechnik}, vol.~63, no.~3, pp. 177--182, 1981.

\bibitem{Cui:10}
T.~Cui, X.~Dong, Z.~Bo, and A.~Juszczyk, ``Hilbert-transform-based
  transient/intermittent earth fault detection in noneffectively grounded
  distribution systems,'' \emph{IEEE Trans. Power Del.}, vol.~26, no.~1, pp.
  143--151, 2011.

\bibitem{Landau:81}
L.~D. Landau and E.~M. Lifshitz, \emph{Quantum Mechanics: Non-relativistic
  Theory}, 3rd~ed.\hskip 1em plus 0.5em minus 0.4em\relax Oxford, UK: Elsevier
  Butterworth-Heinemann, 1981.

\bibitem{Halmos:74}
P.~R. Halmos, \emph{A Hilbert Space Problem Book}.\hskip 1em plus 0.5em minus
  0.4em\relax New York, NY: Springer-Verlag, 1974.

\end{thebibliography}

\end{document}